\documentclass[prl,a4paper,10pt,twocolumn,floatfix,superscriptaddress,amsmath,amsfonts,amssymb,preprintnumbers,citeautoscript,longbibliography]{revtex4-2}

\pdfoutput=1
\usepackage[T1]{fontenc}\usepackage[utf8]{inputenc}
\usepackage{dcolumn,graphicx,color,booktabs,microtype,afterpage} 
\usepackage{sidecap}
\usepackage{times}
\usepackage[colorlinks,plainpages=false,linkcolor=blue,urlcolor=blue,citecolor=blue,pdfpagemode=UseNone,pdfstartview=FitBH]{hyperref}

\usepackage{adjustbox}
\usepackage{amssymb}
\usepackage{gensymb}
\usepackage{amsmath}
\usepackage{graphicx}
\usepackage{braket}
\usepackage{tabularx}
\newcommand{\bygo}{BiYbGeO$_5$}
\newcommand{\biion}{Bi$^{3+}$}
\newcommand{\ybion}{Yb$^{3+}$}
\newcommand{\ybsio}{Yb$_2$Si$_2$O$_7$}

\begin{document}
%
%
%
%
\title
{Localized Triplons and Site Stuffing in the Quantum Dimer Magnet \bygo}
\author{Rachit Kapoor}
\email{ls23aj@brocku.ca}
\affiliation{Department of Physics, Brock University, St. Catharines, ON L2S 3A1, Canada.}
\author{J. Ramirez Diaz}
\affiliation{Department of Physics, Brock University, St. Catharines, ON L2S 3A1, Canada.}
\author{D. R. Yahne}
\affiliation{Neutron Scattering Division, Oak Ridge National Laboratory, Oak Ridge, TN 37831, USA.}
\author{V. O. Garlea}
\affiliation{Neutron Scattering Division, Oak Ridge National Laboratory, Oak Ridge, TN 37831, USA.}
\author{G. Hester}
\email{ghester@brocku.ca}
\affiliation{Department of Physics, Brock University, St. Catharines, ON L2S 3A1, Canada.}
\date{\today}
%
%
%
%
\begin{abstract}
Thermodynamic and muon spin relaxation measurements have recently highlighted \bygo\ as a new example of a rare-earth-based quantum dimer magnet with isolated \ybion\ spin-1/2 dimers. However, direct spectroscopic evidence of the triplet excitations and measurements of the structural disorder are lacking. In this work, polycrystalline \bygo\ was synthesized using conventional high-temperature solid-state methods and investigated via high-resolution neutron powder diffraction and inelastic neutron scattering. Diffraction measurements down to 58 mK reveal no signatures of magnetic order and indicate that nearly 20\% of \ybion\ sites are replaced by non-magnetic \biion, introducing significant structural disorder. Inelastic neutron scattering shows dispersionless triplon excitations, consistent with localized, non-interacting spin dimers. Fits to the triplet excitation spectrum with an XXZ-type anisotropic exchange give $J_{\textnormal{XX}}$ = 0.100(1) meV and $J_\textnormal{Z} = 0.202(1)$ meV. These findings establish \bygo\ as a structurally disordered but magnetically well-isolated quantum dimer system, providing a model platform for studying the resilience of entangled spin states to site dilution.

\end{abstract}

\maketitle
%
%
%
\section{Introduction}
Many exotic quantum states in condensed matter systems, such as quantum spin liquids, emerge from the entanglement of electron pairs into spin singlets. In a quantum spin liquid, the ground state can be described as a superposition of all possible singlet coverings of the lattice \cite{balents_spin_2010}. Understanding how singlets form and respond to perturbations is therefore essential for identifying and designing new quantum spin liquid candidates. 

In cases where long-range entanglement is absent but local singlets form between neighboring spins without breaking lattice symmetries, the resulting state is known as a quantum dimer magnet. These systems are characterized by the absence of long-range magnetic order down to the lowest temperatures and the presence of coherent spin excitations, called triplons. In idealized models with negligible interdimer interactions, the singlet-triplet gap can be closed by an external magnetic field, leading directly to a fully polarized paramagnet. However, real materials invariably host finite interdimer interactions that can stabilize an intermediate field-induced phase, which maps onto a Bose-Einstein condensate of triplons \cite{zapf_bose-einstein_2014}. Quantum dimer magnets thus serve as a minimal platform for exploring quantum entanglement and the onset of quantum criticality in low-dimensional spin systems.

\begin{figure}[!h]
\includegraphics[width = 1.0\columnwidth]{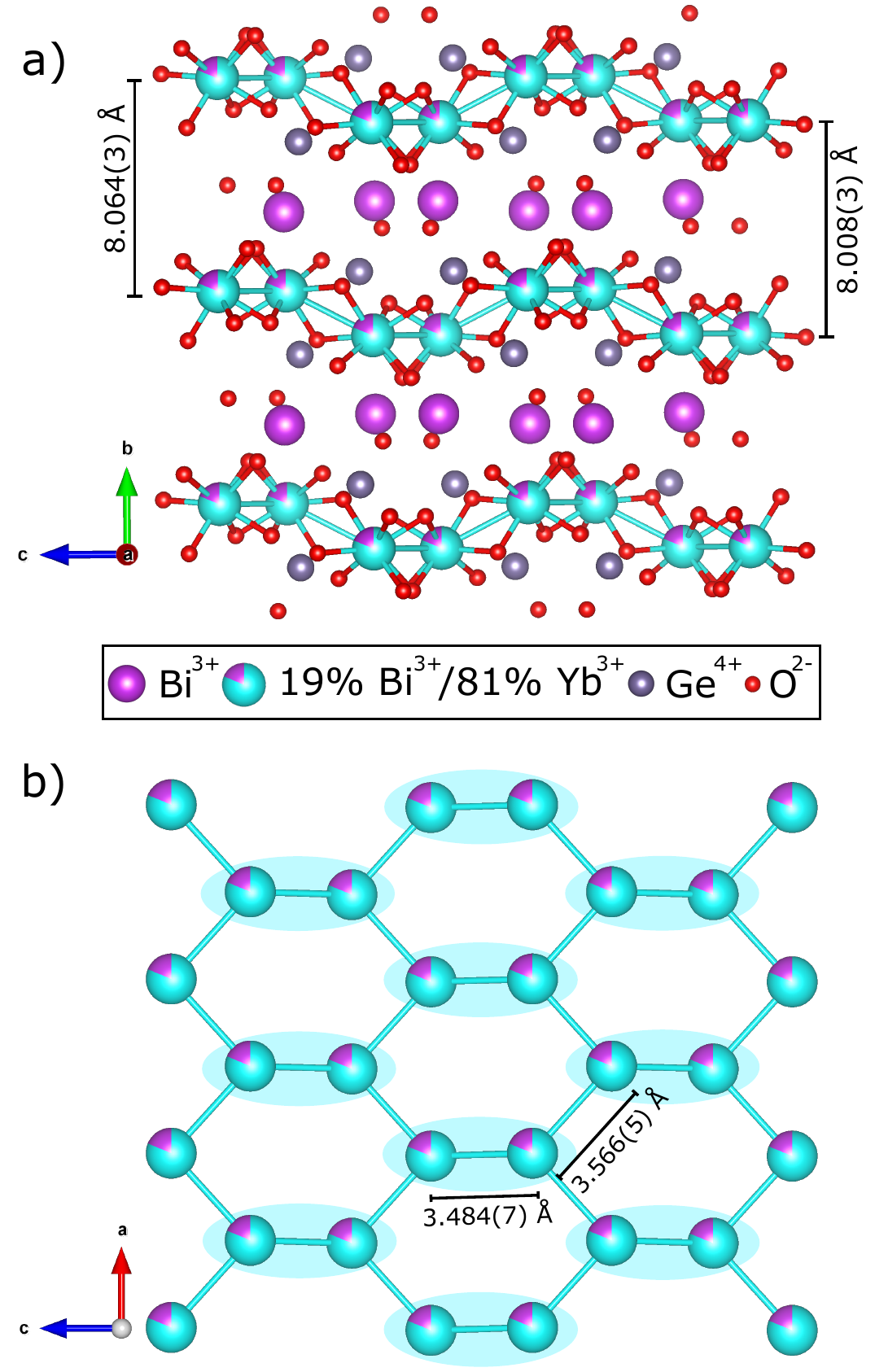}
\caption{a) Crystal structure of \bygo\, viewed along the $a$-axis \cite{momma_vesta_2011}. The interlayer separation alternates symmetrically between 8.064(3) \AA and 8.008(3) \AA, which is considerably larger than the intradimer bond length of 3.484(7) \AA. The fraction of Yb spheres shaded in the Bi color corresponds to the proposed amount ($19.2(26)\%$) of \biion\ stuffing at the \ybion\ site. b) The distorted honeycomb layers are viewed along the $ac$-plane with distortion along the $b$-axis. The dimers are labeled with blue ovals.} \label{fig_crystStruc}
\end{figure}

To date, most quantum dimer magnets have been discovered in 3$d$ transition metal-based compounds, where strong exchange interactions often result in prohibitively high magnetic field strengths (> 20 T) to close the singlet-triplet gap. The discovery of \ybsio\ in 2019 \cite{hester_novel_2019} marked the first rare-earth-based example of such a state, where the contracted nature of the 4$f$ orbitals leads to a relatively small magnetic field to close the singlet-triplet gap ($\sim$1.0 T). Since the discovery of \ybsio\, several other 4$f$-based quantum magnets with emergent singlet ground states have been proposed in Yb$_{2}$SiO$_{5}$ \cite{hase_spin-singlet_2025}, Yb-doped YAlO$_{3}$ \cite{nikitin_experimental_2020}, Yb$_{2}$Be$_{2}$SiO$_{7}$ \cite{brassington_novel_2025}, and BiYbGeO$_{5}$ \cite{mohanty_disordered_2023}. 

\bygo\ is of particular interest as magnetometry, heat capacity, and muon spin relaxation measurements suggest the formation of isolated spin singlets composed of effective spin-1/2 Yb$^{3+}$ moments \cite{mohanty_disordered_2023}. Field-dependent heat capacity measurements by Mohanty \textit{et al.} indicate an XXZ-type anisotropic intradimer exchange with $J_{\textnormal{XX}}$ $\simeq$ 1.3 K and J$_{\textnormal{Z}}$ $\simeq$ 2.6 K with no thermodynamic evidence for interdimer interactions. These measurements confirm the absence of magnetic order at low temperatures and the presence of ground state singlets; however, the ideal method for confirming the presence of triplon excitations -- and their energy scale -- is inelastic neutron scattering, which has not been undertaken until this study.

In this work, we investigate the proposed dimer ground state in \bygo\ using inelastic neutron scattering to search for coherent triplon excitations and assess the degree of Bi$^{3+}$/Yb$^{3+}$ site disorder through high-resolution neutron diffraction. These measurements allow us to probe the strength of the intradimer and interdimer exchange and investigate how the triplon excitations survive in the presence of disorder, a key question for understanding the robustness of dimer physics beyond the clean limit.

\begin{figure*}[!t]
\includegraphics[width = 2.0\columnwidth]{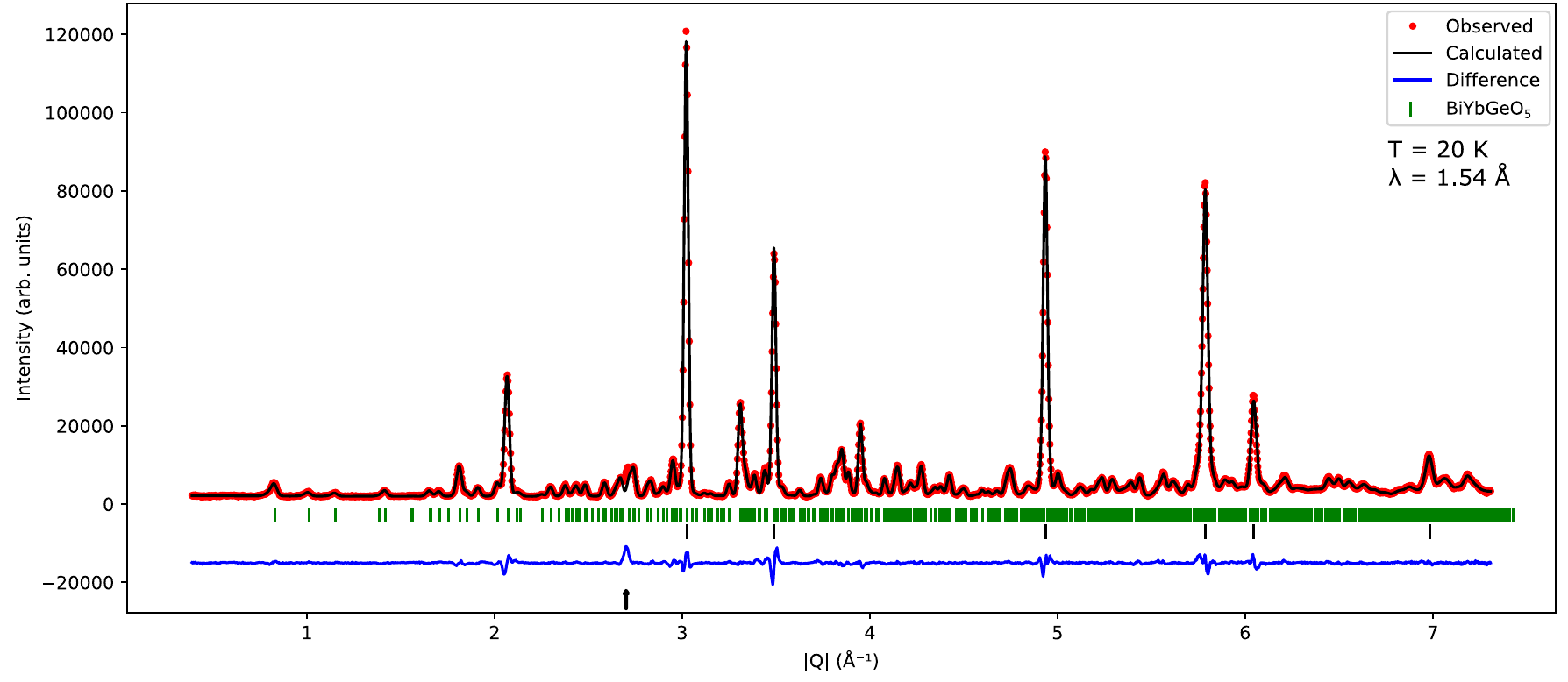}
\caption{A Rietveld refinement of the structure of \bygo\ obtained from neutron diffraction performed at 20 K using an incident wavelength of $\lambda$ = 1.54 \AA. The peaks marked by black ticks come from the Cu canister used for the measurement. These Cu peaks were fitted using Le Bail extraction, not to influence the refinement of the \bygo\ parameters.  A spurious peak visible at $\sim$ 2.71 \AA$^{-1}$ (marked by a black arrow) corresponds to aluminum present in the sample environment due to inadequate collimation at 1.54 \AA\ incident wavelength. The overall $\chi^2$ for the fit is 7.237.} 
 \label{fig_rietveld_refinement}
\end{figure*}


\section{Experimental Methods}
\subsection{Synthesis}
A pale-yellow polycrystalline powder of \bygo\ was synthesized using the solid-state synthesis method. This method is an adaptation from the synthesis method proposed by Cascales \textit{et al.} \cite{cascales_new_2002}. Bi$_2$O$_3$ (99.99\%, Thermo Scientific), Yb$_2$O$_3$ (99.9\%, Sigma-Aldrich), and GeO$_2$ ($\geq$99.99\%, Sigma-Aldrich) were dried to remove any water of crystallization, after which they were stored in a desiccator. Stoichiometric amounts of the dried reactants were taken in molar ratios of $1 : 1 : 2$, respectively, in an agate mortar and pestle and ground, pressed into pellets, heated in air in an alumina crucible inside a box furnace for 24 hours each, and quenched \footnote{It was found with further synthesis that quenching is not necessary and the pellets can be cooled at a slower rate as well}. This process was repeated four times, where the temperature was set to $800~\degree$C, $900~\degree$C, $950~\degree$C, and $950~\degree$C for each heating, respectively. For the last two heatings at $950~\degree$C, the pellets were pressed by adding 1 ml of PVA (polyvinyl alcohol) solution (50 mg/ml) to facilitate proper pellet formation. Phase purity was assessed using powder X-ray diffraction (XRD) at room temperature using a Rigaku SmartLab X-ray Diffractometer ( CuK$_\alpha$ source with K$_{\alpha1}$/K$_{\alpha2}$ of 1.540510\AA/1.544330\AA).

\subsection{Neutron Powder Diffraction}
Neutron diffraction measurements were performed on powdered \bygo\ using the HB-2A Powder Diffractometer (POWGEN) at the Spallation Neutron Source, Oak Ridge National Laboratory, and analyzed using the Rietveld refinement method within the \textsc{FullProf Suite} \cite{rodriguez1993recent}. Approximately 5~g of the compound was loaded in a copper canister backfilled with 10 atm of He to ensure proper thermalization at low temperatures. Data were collected at 20 K, 0.9 K, and 0.058 K. Measurements were taken using two incident wavelengths of $\lambda$ = 2.41 \AA\ and $\lambda$ = 1.54 \AA\ using the Ge(113) and Ge(115) monochromators, respectively. For all measurements, the collimator setup was open-open-12'. A measurement with $\lambda$ = 1.54 \AA\ provided a larger Q range, allowing for more precise measurement of potential site mixing between \biion\ and \ybion.

\subsection{Inelastic Neutron Scattering}

Inelastic neutron scattering (INS) spectra were measured at the Spallation Neutron Source at Oak Ridge National Laboratory using the BL-14B Hybrid Spectrometer (HYSPEC). Slices and cuts of the data were made using the \textsc{Mantid} analysis software \cite{ARNOLD2014156}. Data was collected at low temperature (0.250 K) to observe triplet excitations and at high temperature (100 K) as a background measurement. Multiple datasets were also collected at intermediary temperatures (0.580 K, 0.850 K, 1.5 K, 2.5 K, 10 K, and 20 K) to analyze the evolution of the triplet excitation intensity with increasing temperature. All measurements were performed under no external magnetic field in the unpolarized mode. The Fermi chopper speed was set at 420 Hz, and the lowest incident neutron energy of 3.8 meV was chosen to give the best possible resolution of $\Delta E$ $\simeq$ 0.1 meV at the elastic line. All data, except for 0.250 K, were collected using a detector bank covering a scattering angle (2$\theta$) range of $8\degree - 68\degree$, giving horizontal detector coverage of 0.20 \AA$^{-1}$ to 1.5 \AA$^{-1}$ at the elastic line. To better measure the structure factor of the dimers at 250 mK, an extended detector coverage of $8\degree - 95\degree$ was used to give a |Q| range of 0.20 \AA$^{-1}$ to 2.0 \AA$^{-1}$ at the elastic line.


\section{Results} 

\begin{figure*}[t]
\includegraphics[width = 2.0\columnwidth]{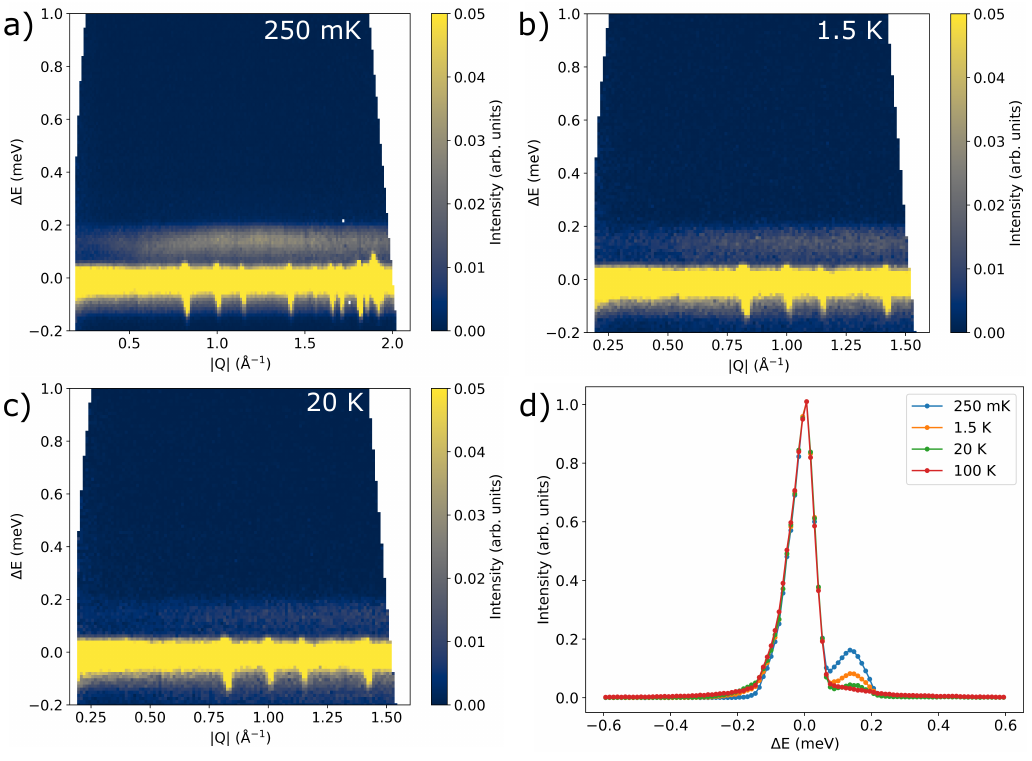}
\caption{a-c) $\Delta$E vs |Q| color contour slices of INS data at a) 250 mK, b) 1.5 K, and c) 20 K showing a flat triplon mode visible above the bright elastic line. The scale factor for all panels has been kept the same to show the triplon band getting fainter on increasing temperature, signifying the depopulation of the ground state. d) Cuts of the INS data made along $\Delta$E integrating |Q| from 0.25 \AA$^{-1}$ to 1.4 \AA$^{-1}$ for 250 mK, 1.5 K, 20 K, and 100 K. The intensities of all scans have been normalized to allow comparison between the spectral weights of the triplon peak as a function of temperature.The large peak at 0.0 meV is the elastic line. A temperature-dependent triplon mode centered at 0.135 meV is visible for all three temperatures. \label{hyspec_slices}} 
\end{figure*}

\subsection{Neutron Powder Diffraction}

Refinement of the 20 K dataset yielded lattice parameters of \textit{a} = 5.2866(1) \AA, \textit{b} = 15.1762(3) \AA, and \textit{c} = 10.9333(2) \AA, smaller than previously reported room-temperature values by Mohanty \textit{et al.} \cite{mohanty_disordered_2023} and Cascales \textit{et al.} \cite{cascales_new_2002} (see Figure~\ref{fig_crystStruc}). This contraction of the unit cell upon cooling is consistent with typical thermal contraction behavior.

Structural refinements of the $\lambda$ = 1.54 \AA\ scan at 20 K, further revealed a degree of site disorder between the \biion\ and \ybion\ ions, likely driven by the similar ionic radii of \biion\ (1.17 \AA) and \ybion\ (0.985 \AA) \cite{shannon_revised_1976}, their identical valence states, and that all atoms occupy the same Wyckoff position, imposing identical symmetry constraints. Approximately 19.2(26)\% of the \ybion\ (occupying the Yb site, see Table~\ref{table_refinement}) are found to be replaced by \biion, as determined by Rietveld analysis. Alternative scenarios, including \ybion\ occupancy of \biion\ sites and bidirectional site mixing, were tested but yielded unrealistic parameters in the form of near-zero occupancy values, showing that the refinement does not support these site-mixing cases. Given that \biion\ is non-magnetic, this substitution effectively breaks Yb-Yb magnetic dimers, generating an equivalent fraction of unpaired \ybion\ spins. Such site mixing defects are expected to locally increase the average interdimer distances and may play a crucial role in the low-temperature magnetic behavior of \bygo.\

The absence of additional Bragg reflections across the entire measured temperature range confirms that all peaks are structural in origin, indicating a lack of long-range magnetic order down to 58 mK. To probe the possibility of short-range magnetic correlations, the diffraction pattern measured at 20 K was subtracted from the 0.058 K pattern. The resulting difference was featureless within the noise level, indicating the absence of any diffuse scattering and ruling out the presence of incipient magnetic transitions or short-range order. 

Two temperature-independent impurity peaks are visible at $Q \approx 1.93$ \AA$^{-1}$ and $2.11$ \AA$^{-1}$ with very small intensities. These peaks are not visible in the shorter-wavelength ($\lambda$ = 1.54 \AA) scan shown in Figure~\ref{fig_rietveld_refinement}, likely due to reduced resolution at higher |Q| values where peak overlap becomes more significant. The impurity peaks are visible in the longer wavelength neutron scans and the XRD scan, which have been provided in the supplemental information (SI) \cite{SI}.

\renewcommand{\arraystretch}{1.5} 

\begin{table}[htbp]
\centering
\begin{adjustbox}{width=\textwidth/2}
\begin{tabular}{ccccccc}
\hline
\hline
\textbf{Atom} & \textbf{Site} & \textit{x} & \textit{y} & \textit{z} & \textbf{B$_{\textnormal{iso}}$} & \textbf{Occupancy} \\
\hline
Bi1 & 8c & 0.95685(69) & 0.23741(24) & 0.14649(38) & 0.130(66) & 1 \\
Yb & 8c & 0.00808(65) & 0.05396(15) & 0.35944(30) & 0.042(50) & 0.808(26) \\
Bi2 & 8c & 0.00808(65) & 0.05396(15) & 0.35944(30) & 0.042(50) & 0.192(26) \\
Ge & 8c & 0.00543(93) & 0.40456(27) & 0.40269(34) & 0.282(62) & 1 \\
O1 & 8c & 0.07054(97) & 0.30183(32) & 0.32959(48) & 0.21(11) & 1 \\
O2 & 8c & 0.3002(11) & 0.43740(39) & 0.45670(53) & 0.19(8) & 1 \\
O3 & 8c & 0.3182(14) & 0.12552(40) & 0.47399(56) & 0.62(13) & 1 \\
O4 & 8c & 0.2542(13) & 0.15041(43) & 0.23658(57) & 0.50(10) & 1 \\
O5 & 8c & 0.3493(11) & 0.47814(43) & 0.19261(54) & 0.37(10) & 1 \\
\hline
\hline
\end{tabular}
\end{adjustbox}
\caption{The refined parameters of atomic coordinates \textit{(x, y, z)}, isotropic atomic displacement \textit{(B$_\textnormal{iso}$)}, and site occupancy obtained by the Rietveld refinement of the neutron diffraction pattern of \bygo\ taken at 20 K using $\lambda$ = 1.54 \AA\ incident wavelength.\label{table_refinement}}
\end{table}

\subsection{Inelastic Neutron Scattering}

To investigate the low-energy spin dynamics in \bygo, INS measurements were carried out using HYSPEC. As shown in Figures~\ref{hyspec_slices}a-c, a well-defined excitation band is observed across a broad range of momentum transfers, |Q|. This excitation appears dispersionless within the instrumental resolution of $\sim$ 0.1 meV, consistent with isolated dimers with negligible interdimer exchange. Energy transfers were measured up to 3.61 meV (see SI \cite{SI}), in which no crystalline electric field (CEF) excitations were detected, placing a lower bound on the CEF gap of 42 K.

The normalized scattering intensity was plotted as a function of energy transfer ($\Delta$E) by integrating a |Q| range of 0.25 \AA$^{-1}$ -- 1.4 \AA$^{-1}$ as shown in Figure~\ref{hyspec_slices}d. Although the triplon excitation energy remains temperature-independent, its spectral weight decreases with increasing temperature, consistent with thermal depopulation of the singlet ground state.

\begin{figure}[!h]
\includegraphics[width = 1.0\columnwidth]{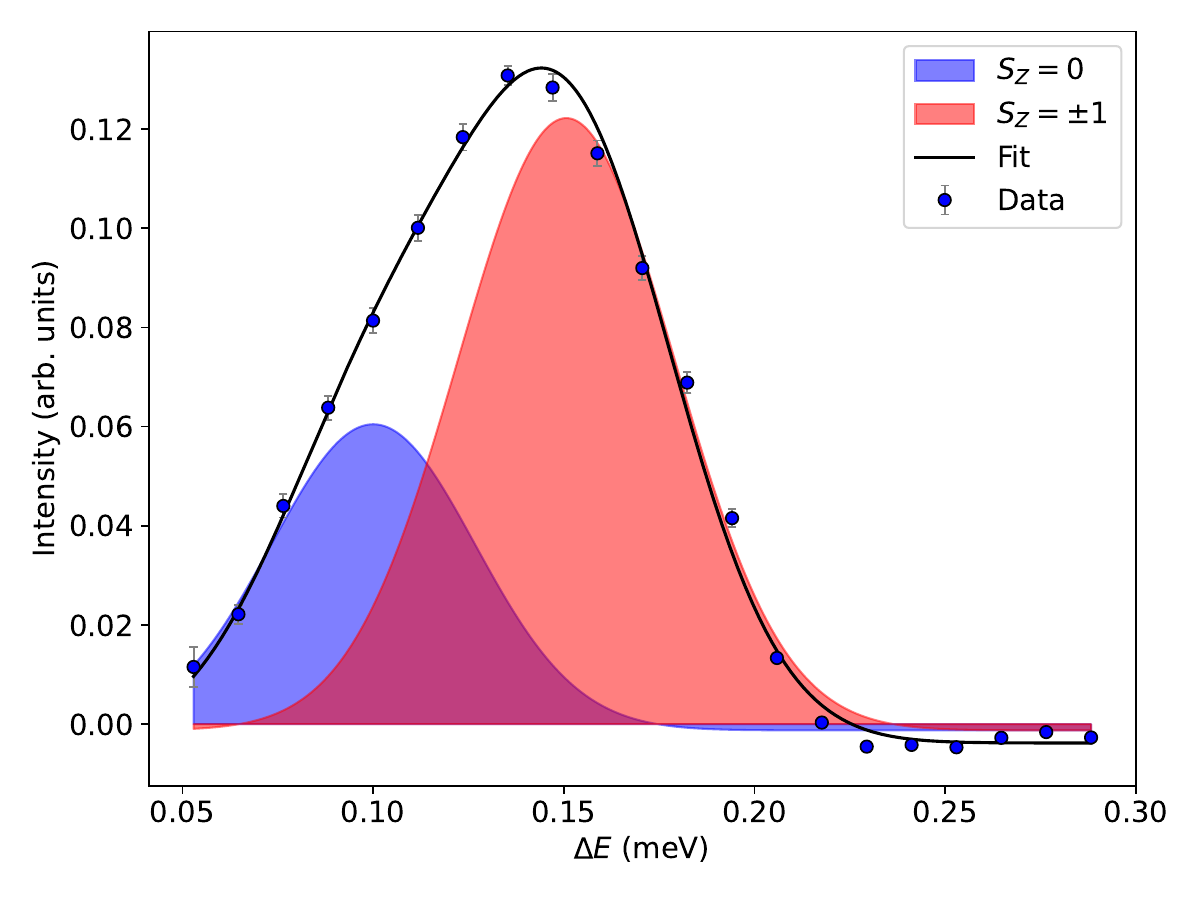}
\caption{A cut along $\Delta$E made by subtracting the 100 K INS data from the 250 mK (integrating |Q| from 0.25 \AA\ to 1.4 \AA). This excitation has been fit to the sum of two Gaussians, centered at 0.100(1) meV and 0.1510(4) meV, consistent with an XXZ-type intradimer exchange interaction. Using the Hamiltonian for an XXZ isolated dimer magnet, the exchange parameters come out to be 0.100(1) meV and 0.202(1) meV. The full widths at half maximum for the Gaussians are 0.06 meV and 0.07 meV, respectively, which are smaller than the resolution of HYSPEC ($\sim$ 0.1 meV).} \label{gaussian_fit}
\end{figure}

To isolate the triplon excitation, the high-temperature (100 K) data were subtracted from the low-temperature (0.250 K) dataset, effectively removing the elastic background and sample environment. The resulting difference spectra show a single excitation that is limited by the instrument resolution. A two-Gaussian model was considered to estimate the intradimer exchange interactions, consistent with the expectations of an XXZ-type anisotropic exchange in isolated dimers as proposed by heat capacity and magnetometry calculations done by Mohanty \textit{et al.} \cite{mohanty_disordered_2023}. Since the synthesis method used in this work is different in some respects from that of Mohanty \textit{et al.}, single and triple-Gaussian models were also tested, resulting in comparable agreements with the INS data. The single and triple-Gaussian fits, with their fitted parameters, have been included in the SI \cite{SI}. In Figure~\ref{gaussian_fit}, we have presented the double peak model because it is consistent with the XXZ energy hierarchy proposed by Mohanty \textit{et al.} In order to more precisely determine the exchange interactions, a higher-resolution INS scattering experiment would be required.

In the XXZ dimer framework at zero magnetic field, three triplet excitations are expected: a single $S_z = 0$ state and a doubly degenerate excitation to the $S_z = \pm 1$ states, with the latter state expected to have double the spectral weight of the former. Under this assumption, the intensities of the two peaks were fixed in the ratio 1:2 and the extracted exchange parameters were found to be $J_{\textnormal{XX}} = 0.100(1)$~meV and $J_{\textnormal{Z}} = 0.202(1)$~meV, yielding an anisotropy ratio of $J_{\textnormal{XX}}/J_{\textnormal{Z}} \approx 0.5$. To avoid the elastic line from affecting the fit to the triplon peak, only data above 0.053 were used for the fit. The energy hierarchy agrees qualitatively with that proposed by Mohanty \textit{et al.} \cite{mohanty_disordered_2023}. This hierarchy was determined by Mohanty \textit{et al.} using heat capacity measurements, which are especially sensitive to energy hierarchies. While the $J_{\textnormal{XX}}$ and $J_{\textnormal{Z}}$ values differ by $\sim$ 0.01 meV and $\sim$ 0.02 meV, respectively, this can be attributed to the differences caused by indirect means of extracting the anisostropic exchange interactions using Gaussian fitting and simulations of heat capacity data by this work and Mohanty \textit{et al.}, respectively (see Table~\ref{exchange_interactions}). The average intradimer coupling, on the other hand, can be very clearly determined from the high-resolution INS spectra to be 0.1354(3) meV, which is different from $\approx 0.15$ meV extracted using heat capacity modeling. This discrepancy emphasizes the utility of direct spectroscopic probes such as INS for resolving anisotropic exchange strengths and may reflect inherent limitations of thermodynamic fitting approaches.

\begin{table}[htbp]
\centering
\begin{adjustbox}{width=\textwidth/2}
\begin{tabular}{ccc}
\hline
\hline
\centering
\textbf{Exchange interactions} & \textbf{This Work (meV)} & \textbf{Mohanty \textit{et al.} (meV)}\\
\hline
$J_{\textnormal{XX}}$ & 0.100(1) & $\simeq$ 0.11 \\
$J_{\textnormal{Z}}$ & 0.202(1) & $\simeq$ 0.22\\
\hline
\hline
\end{tabular}
\end{adjustbox}
\caption{A comparison between the anisotropic exchange parameters measured in this work with INS and by Mohanty \textit{et al.} by fitting heat capacity data \cite{mohanty_disordered_2023}.}
\label{exchange_interactions}
\end{table}

\begin{figure}[h]
\includegraphics[width = 1.0\columnwidth]{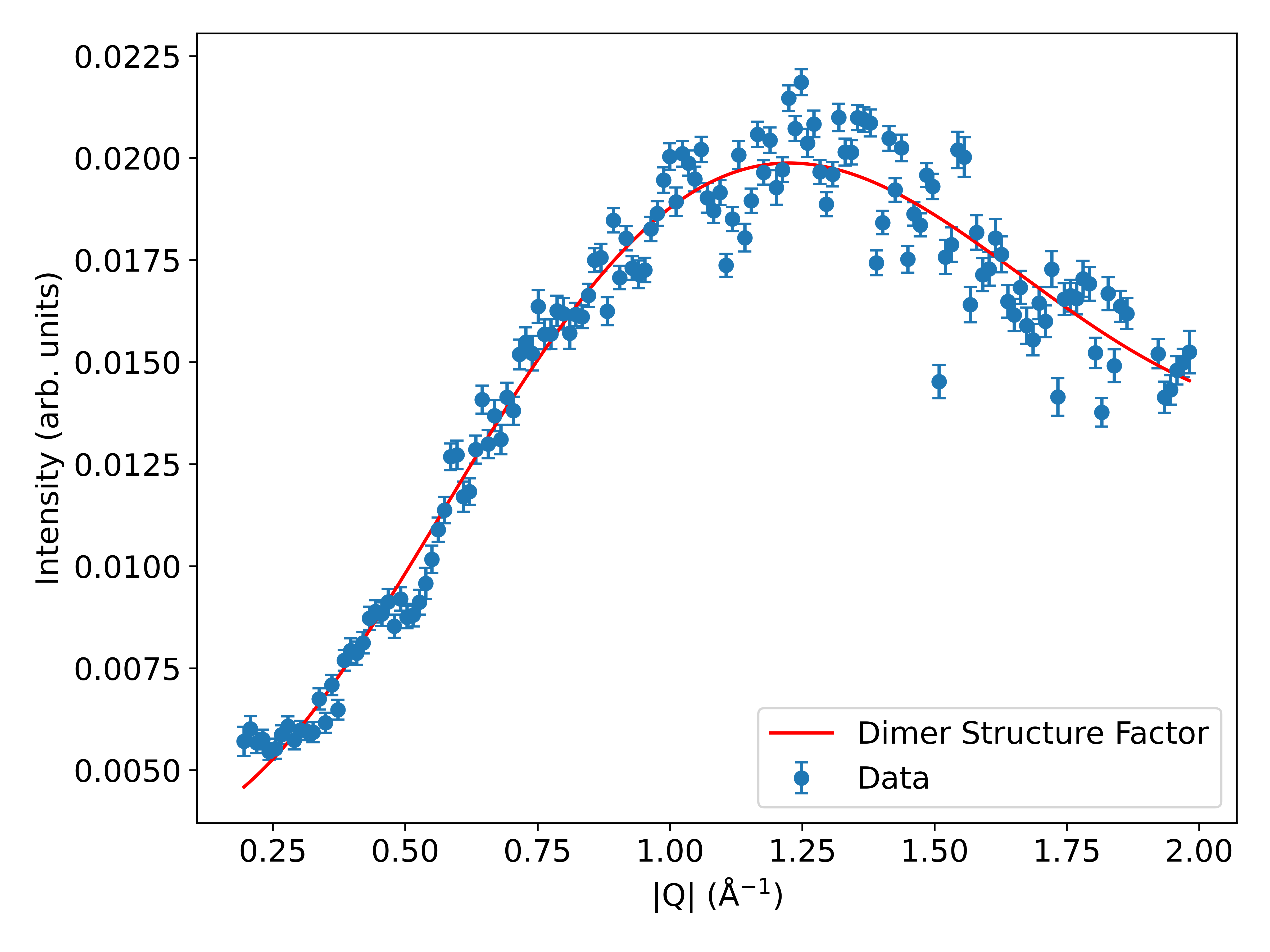}
\caption{A cut of the INS data taken at 250 mK made along |Q| by integrating $\Delta$E symmetrically around the 0.135 meV maximum from 0.070 meV to 0.200 meV. Four data points between 1.87\AA\ and 1.92 \AA\ were removed from the fit as they were present due to the elastic tail extending into the triplon band. The fitted line represents the calculated structure factor of a dimer magnet system that yields an intradimer distance of 3.541(11) \AA.} \label{structure_factor_fit}
\end{figure}

The $Q$-dependence of the triplet excitation at 0.250 K was further analyzed to confirm the spatial configuration of Yb dimers as shown in Figure~\ref{structure_factor_fit}. The measured intensity was fit to the magnetic structure factor for an isolated dimer,
$$I(Q) = r_0^2\frac{k'}{k}Np\times(f(Q))^2\times \left( 1-\frac{sin(Qd)}{Qd}\right)$$
where, $f(Q)$ is the magnetic form factor for \ybion\ and $d$ is the intradimer distance. The magnetic form factor was modeled as $f(Q)=j_0 + \left(1-\frac{g_S}{g}\right)j_2$ \cite{lovesey_theory_1986}, using tabulated coefficients for \ybion\ \cite{freeman_dirac-fock_1979}. The dimer distance has been extracted to be 3.541(11) \AA. However, it is difficult to make definitive statements about possible dimerization involving next-nearest-neighbor \ybion\ ions in the presence of broken nearest-neighbor dimers, since the nearest-neighbor (3.494(9) \AA) and next-nearest-neighbor (3.557(6) \AA) distances are too close to be distinguished within the resolution of the INS data (see SI). Future single-crystal INS measurements, with improved Q-resolution, could provide a more sensitive probe of whether alternative dimer configurations emerge in the presence of site disorder.

\section{Discussion}

\bygo\  exhibits signatures of an isolated quantum dimer magnet, with a significant degree of Bi–Yb site mixing ($\simeq$ 19.2\%) leading to broken Yb-Yb dimers. The substitution of non-magnetic \biion\ for \ybion\ both increases the average separation between remaining magnetic dimers and may play a role in further suppressing interdimer triplon hopping, accounting for the flat band observed in inelastic neutron scattering. This level of site mixing occurs despite stoichiometric synthesis, implying that excess \biion\ must occupy \ybion\ sites and expelled \ybion\  segregates into minor impurity phases. The number and intensity of these impurity peaks are too low to enable precise identification of the impurity phases. No signs of magnetic excitations from the impurity phase are observed in the inelastic neutron scattering data obtained from HYSPEC, likely due to the relatively small amount of impurity.

Previous work by Mohanty \textit{et al.} \cite{mohanty_disordered_2023} and Cascales \textit{et al.} \cite{cascales_new_2002} utilized X-ray diffraction data of \bygo\ to assess phase purity and therefore they do not report any Yb/Bi site mixing. However, the X-ray cross-sections of \biion\ ($2.116\times10^3$ barns) and \ybion\ ($1.454\times10^3$ barns) are fairly close, as that is predominantly a function of the number of electrons in the ions \cite{1570572700268409088}. The neutron cross-sections, by contrast, are quite distinct (9.315 barns for Bi and 19.42 barns for Yb), allowing one to discern the site mixing happening between the two ions \cite{varley1992neutron}. The presence of site mixing is perhaps unsurprising as Cascales \textit{et al.} also analyzed the XRD pattern of BiYGeO$_5$, which showed 12\% of \biion\ stuffing in the Y$^{3+}$ site. In the work of Mohanty \textit{et al.}, they report an impurity contribution of $\simeq 4.2\%$ that is required to appropriately fit their field-dependent heat capacity data, assuming the impurity spin to be $S = 1/2$. The presence of this impurity may be explained by \biion\ stuffing into the \ybion\ site where the substitution of a Yb-Yb dimer by a Bi-Yb pair results in an unpaired \ybion\ spin, which could yield this measured "impurity" contribution to magnetic susceptibility.

Site mixing effects of this sort are well known in Yb-based magnets and can dramatically alter magnetic behavior. For example, the pyrochlore quantum spin ice Yb$_2$Ti$_2$O$_7$ is notoriously sensitive to small off-stoichiometry. Even $\sim$2\% “stuffing” of the Ti$^{4+}$ sites by extra Yb$^{3+}$ leads to strongly sample-dependent ground states \cite{gaudet_neutron_2015, ghosh_effects_2018}. Stoichiometric polycrystalline Yb$_2$Ti$_2$O$_7$ shows a sharp transition, whereas lightly stuffed single crystals exhibit broad heat-capacity anomalies and no conventional order \cite{ross_lightly_2012}; demonstrating that magnetic properties can differ significantly between single crystals and polycrystalline samples due to subtle synthesis-dependent variations in disorder. By analogy, \bygo’s site mixing similarly introduces random disruption, but in this case, the disorder further suppresses interdimer triplon hopping rather than creating competing couplings. Future growth of \bygo\ single crystals will be crucial to determine whether this site mixing is an intrinsic feature of the material or a consequence of polycrystalline synthesis.

Another instructive comparison is the Shastry–Sutherland, isolated dimer system Yb$_2$Be$_2$SiO$_7$. In this system, the site mixing between the non-magnetic Be$^{2+}$ and Si$^{4+}$ ions leads to a significant broadening of the triplon band as observed in INS measurements \cite{brassington_novel_2025}. Er$_2$Be$_2$SiO$_7$, by contrast, does not show the Be/Si site mixing and develops long-range magnetic ordering upon cooling  \cite{brassington2024magnetic}. Although the disorder in Yb$_2$Be$_2$SiO$_7$ is not at the magnetic site, it still perturbs the effective exchange landscape. 

Beyond the experimental implications, the isolated dimers of \bygo\ may be an ideal system for testing quantum simulations of entangled states in condensed matter systems. Quantum dimer magnets serve as an ideal two-qubit prototype for quantum simulation, directly mapping each spin in a dimer to a qubit pair and encoding the antiferromagnetic exchange interaction into entangling gates \cite{ardavan_engineering_2015, jing-min_two-qubit_2005}. Eassa \textit{et al.} demonstrated that resource-efficient fast-forwarding measurements performed with short-depth circuits can faithfully reproduce the dynamical structure factor and the triplon gap of dimers on current quantum processors, with fidelities high enough to match costly INS measurements \cite{eassa_high-fidelity_2024}. \bygo\ may be an ideal system for extending that analysis to rare-earth-based quantum magnets.

\section{Conclusions}
In this work, polycrystalline samples of \bygo\ were synthesized via conventional solid-state reactions, and their magnetic and structural properties were investigated through neutron powder diffraction and inelastic neutron scattering. Neutron diffraction measurements were carried out at 20 K, 0.9 K, and 58 mK, confirming the absence of magnetic orderings. Structural refinements revealed significant site disorder, with 19.2(26)\% of \ybion\ sites replaced by non-magnetic \biion\ ions. Inelastic neutron scattering measurements uncovered dispersionless triplon excitations, consistent with localized triplet modes due to isolated Yb–Yb dimers. Gaussian fits to the energy spectrum yielded an XXZ-type anisotropic exchange interaction, with $J_{\textnormal{XX}} = 0.100(1)$ meV and $J_{\textnormal{Z}} = 0.202(1)$ meV. Analysis of the magnetic structure factor further determined the Yb-Yb dimer distance to be 3.541(11) \AA. These results confirm that despite substantial structural disorder, \bygo\ hosts a magnetically disordered quantum ground state composed of well-isolated Yb dimers, making it a compelling platform for exploring disorder-resilient quantum magnetism.

\section{Acknowledgements}
R. Kapoor and G. Hester would like to thank R. Ganesh for insightful conversations closely and distantly related to this work. We would also like to thank C. Sarkis for facilitating the measurements using HYSPEC for this work. We also extend our gratitude to Carter Fortuna for synthesizing many variants of this material. A portion of this work used resources at the High-Flux Isotope Reactor and the Spallation Neutron Source, which are DOE Office of Science User Facilities operated by Oak Ridge National Laboratory. The beamtime was allocated to POWDER (HB-2A) and HYSPEC (BL-14B) on proposal numbers IPTS-32979 and IPTS-35019, respectively. 

%
%
\nocite{denisova_high-temperature_2017}
\bibliography{bygo}

\end{document}


%
%
%
%
\title
{Localized Triplons and Site-Stuffing in the Quantum Dimer Magnet \bygo}
\section{SUPPLEMENTARY INFORMATION}
\author{Rachit Kapoor}
\email{ls23aj@brocku.ca}
\affiliation{Department of Physics, Brock University, St. Catharines, ON L2S 3A1, Canada.}
\author{J. Ramirez Diaz}
\affiliation{Department of Physics, Brock University, St. Catharines, ON L2S 3A1, Canada.}
\author{D. R. Yahne}
\affiliation{Neutron Scattering Division, Oak Ridge National Laboratory, Oak Ridge, TN 37831, USA.}
\author{V. O. Garlea}
\affiliation{Neutron Scattering Division, Oak Ridge National Laboratory, Oak Ridge, TN 37831, USA.}
\author{G. Hester}
\email{ghester@brocku.ca}
\affiliation{Department of Physics, Brock University, St. Catharines, ON L2S 3A1, Canada.}
\date{\today}
\maketitle

\section{X-RAY DIFFRACTION}

The XRD pattern on our sample used for neutron scattering shows three additional impurity peaks at |$\Vec{Q}$| = 1.86 \AA, 1.93 \AA, and 2.10 \AA. The intensity of the peaks is weak with normalized intensities of 0.04, 0.03, and 0.04, respectively. The peak seen at 1.86 \AA\ is only visible in the XRD pattern but is invisible in all neutron powder diffraction (NPD) scans. The number and intensity of these impurity peaks are too low to enable precise identification of the impurity phase(s).

\begin{figure}[h]
    \centering
    \includegraphics[width=1\linewidth]{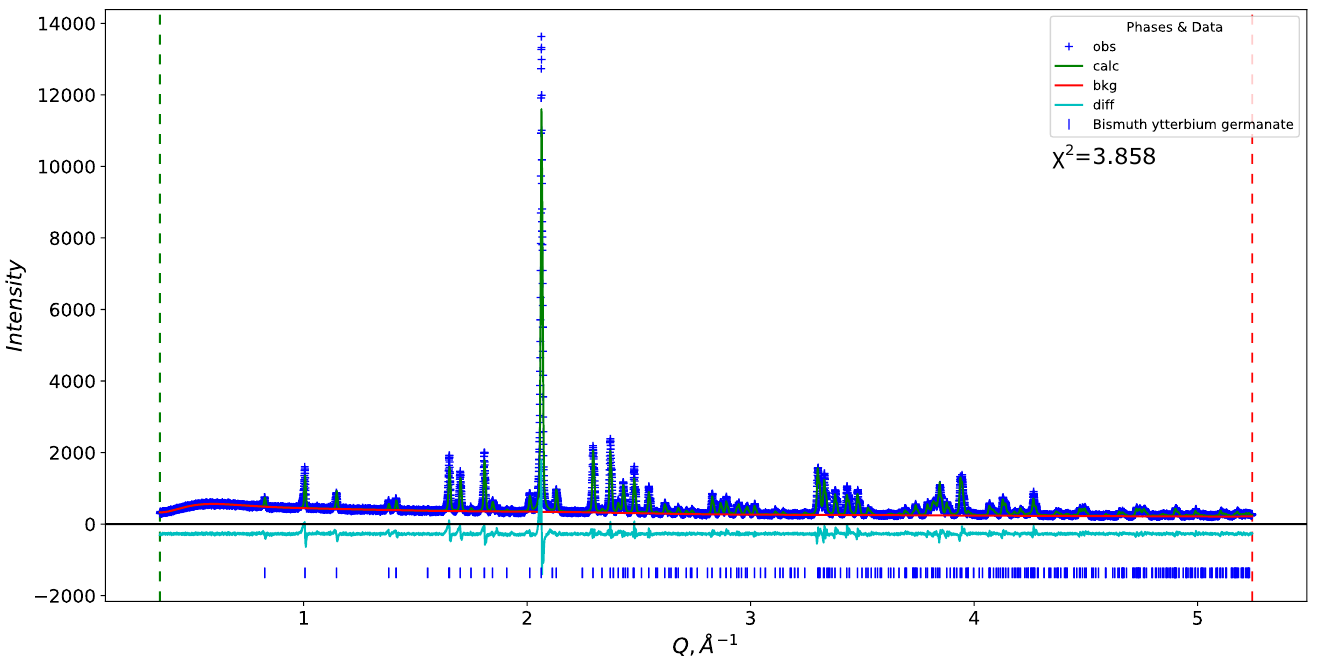}
    \caption{A Rietveld refinement of the structure of \bygo\ obtained from X-ray diffraction performed at room temperature using a Rigaku SmartLab X-ray Diffractometer (CuK$_\alpha$ source with K$_{\alpha1}$/K$_{\alpha2}$ wavelength of 1.540510\AA/1.544330\AA).}
    \label{fig:enter-label}
\end{figure}

\begin{figure}[h]
    \centering
    \includegraphics[width=1\linewidth]{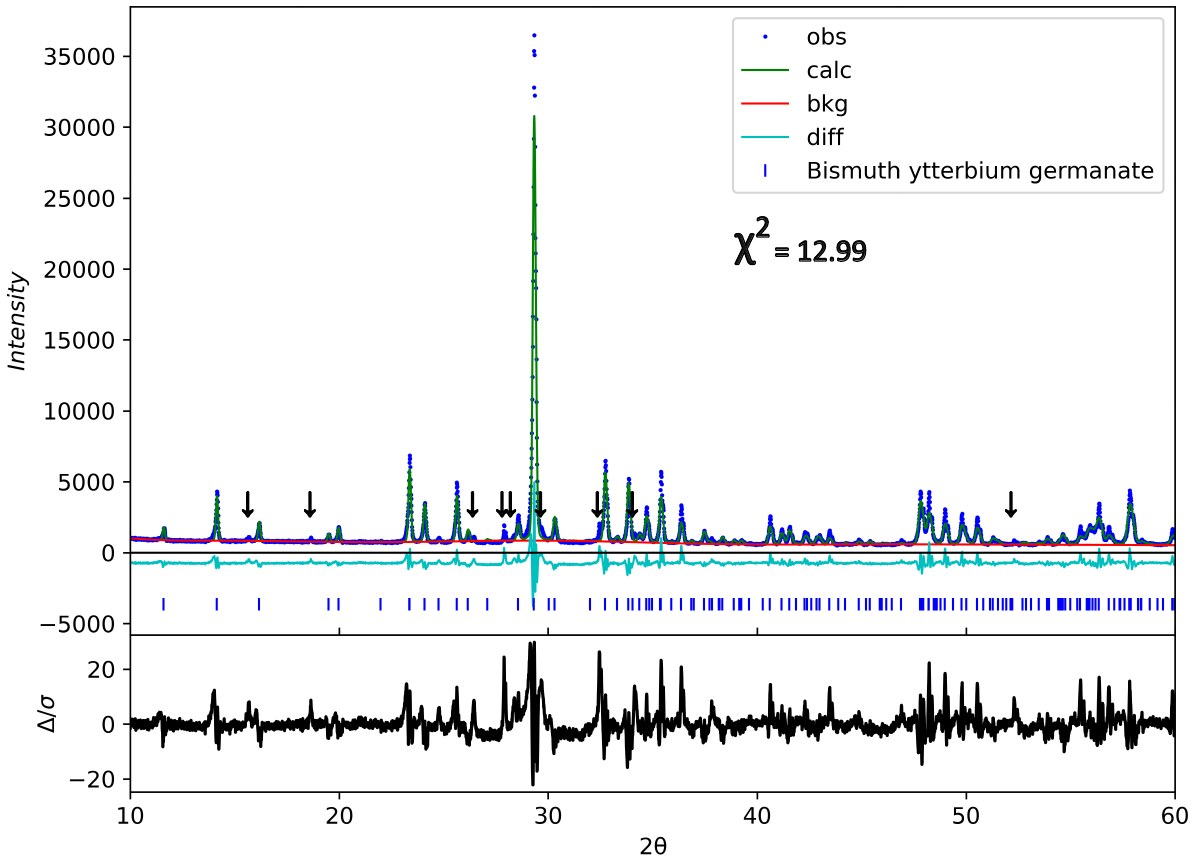}
    \caption{A Rietveld refinement of the structure of \bygo\ following synthesis instructions by Mohanty \textit{et al.} \cite{mohanty_disordered_2023} obtained from X-ray diffraction performed at room temperature using a Rigaku SmartLab X-ray Diffractometer (CuK$_\alpha$ source with K$_{\alpha1}$/K$_{\alpha2}$ wavelength of 1.540510\AA/1.544330\AA). The black arrows mark all of the impurity peaks obtained using this synthesis method.}
    \label{mohanty_synthesis_XRD}
\end{figure}

Mohanty \textit{et al.} \cite{mohanty_disordered_2023} employed a different synthesis route. While they cited Cascales \textit{et al.} \cite{cascales_new_2002} in their paper, they did not follow the procedure as given by Cascales. Mohanty \textit{et al.} report finding no impurities in their sample. Their published XRD pattern makes it hard to see if there are any unaccounted peaks. We reproduced their synthesis method, and found that the compound prepared exactly according to their procedure had a greater number of impurity peaks than the one we prepared for our studies using the method listed by Cascales \textit{et al.} (Figure~\ref{mohanty_synthesis_XRD}). We have included the XRD pattern of the scan done using Mohanty’s procedure, showing impurity peaks at 15.7$\degree$, 18.6$\degree$, 26.4$\degree$, 27.9$\degree$, 28.4$\degree$, 29.7$\degree$, 32.5$\degree$, 34.1$\degree$, and 52.3$\degree$. It is also important to note that we expect that the compound cannot be synthesized as pure in its polycrystalline form due to the observations of the same Bi stuffing into the rare-earth site in the yttrium version of the compound by Cascales \textit{et al. }in 2002. The work of Denisova \textit{et al.} (2017) \cite{denisova_high-temperature_2017} also indicates that BiYbGeO$_5$ forms along the intersection of the phase diagrams of the two independent compounds Yb$_2$GeO$_5$ and Bi$_2$GeO$_5$, suggesting that the similarity in ionic radii and the associated propensity for site mixing may play a role in stabilizing this phase.  \\

However, considering Mohanty \textit{et al.}'s XXZ values are reasonably similar to what we can interpolate from our INS data, they likely have a comparable composition of the desired compound. The system's essential features remain consistent: both studies find evidence for an isolated dimer ground state. In our case, a dispersionless triplon excitation is observed, which agrees with the dimer model proposed by Mohanty \textit{et al.} Furthermore, the lattice parameters obtained from our diffraction refinements closely match those reported previously.

\newpage


\section{NEUTRON POWDER DIFFRACTION}

Neutron diffraction measurements performed at 20 K using 1.54 \AA\ incident wavelength show an unindexed peak visible in the pattern at 2.71 \AA$^{-1}$ corresponding to the aluminum can. This peak is not visible in any of the neutron diffraction patterns using the longer incident wavelengths or the XRD pattern. The other three scans were performed with an incident wavelength of 2.41 \AA. Two additional peaks at $Q \approx 1.93$ \AA$^{-1}$ and $2.11$ \AA$^{-1}$ were detected at all temperatures with very small intensities.

The lattice parameters obtained for each scan type are given in Table~\ref{lattice_parameters}. For the refinement of these patterns, the isotropic atomic displacement (B$_{\textnormal{iso}}$) was refined only for the 1.54 \AA\ scan because the peaks at higher |Q| are more sensitive to this parameter. Since high |Q| data is not included in the $\lambda$ = 2.41 \AA\ data, B$_{\textnormal{iso}}$ values from the 1.54 \AA\ scan were used and kept unrefined for all 2.41 \AA\ scans.

\renewcommand{\arraystretch}{1.5}
\begin{table}[h]
\centering
    \begin{tabular}{cccc}
    \hline
    \hline
         \textbf{Temperature (Incident Wavelength)} & \textbf{a} & \textbf{b} & \textbf{c}\\
         \hline
         20 K (1.54 \AA) & 5.2866(1) & 15.1762(3) & 10.9333(2)\\
        20 K (2.41 \AA) & 5.2851(1) & 15.1720(3) & 10.9254(2)\\
        900 mK (2.41 \AA) & 5.2854(1) & 15.1726(2) & 10.9261(2)\\
        50 mK (2.41 \AA) & 5.2849(1) & 15.1718(3) & 10.9252(2)\\
        \hline
        \hline
    \end{tabular}
    \caption{Lattice parameters \textit{a, b, c} (in \AA) as given by refining different sets of neutron diffraction.}
    \label{lattice_parameters}
\end{table}

\begin{figure}[h]
    \centering
    \includegraphics[width=1\linewidth]{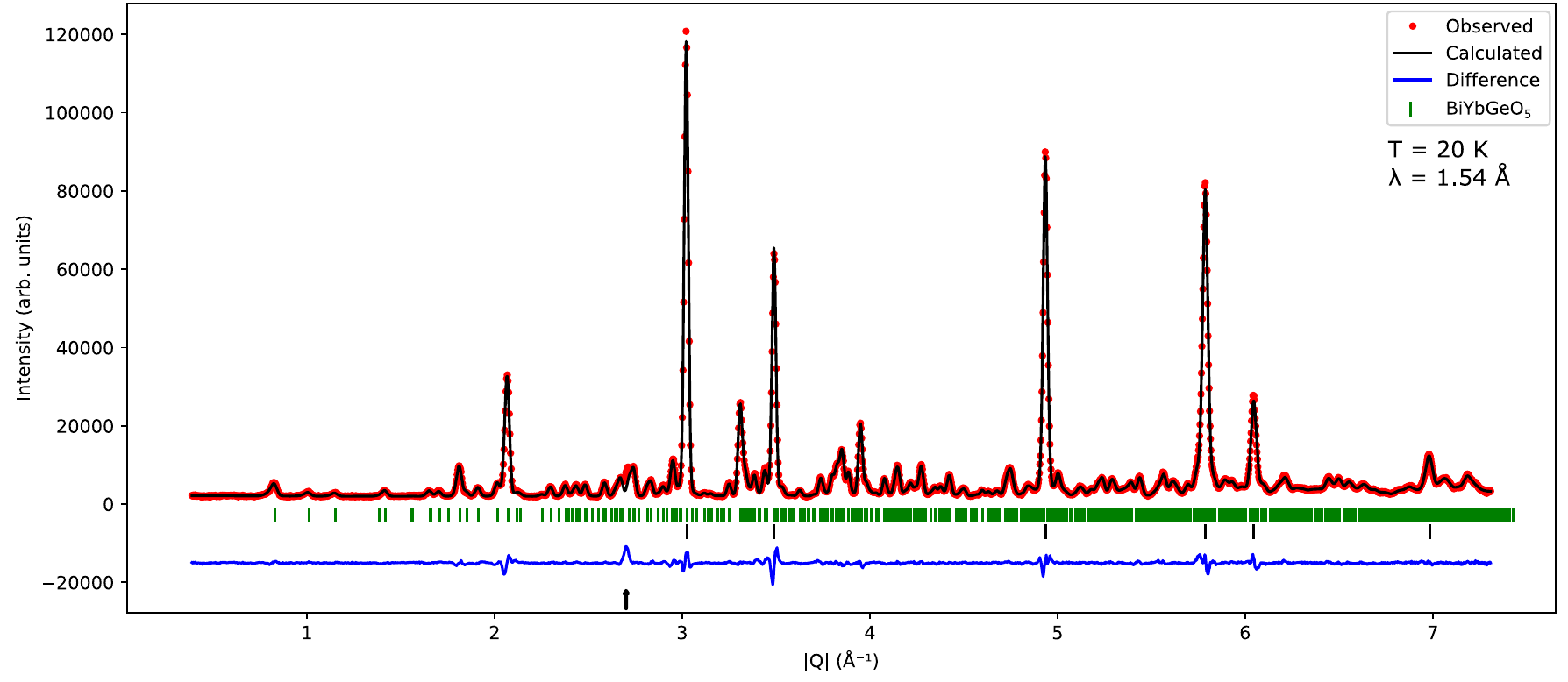}
    \caption{A Rietveld refinement of the structure of \bygo\ obtained from neutron diffraction performed at 20 K using an incident wavelength of $\lambda$ = 1.54 \AA. The peaks, marked by black arrows, come from the Cu canister used for the measurement. These Cu peaks were fitted using Le Bail extraction, not to influence the refinement of the \bygo\ parameters. The \textnormal{$R_{bragg}$} for the pattern from \bygo\ is 3.94 ($\chi^2=7.20$). This is the same as Figure 2 in the main text and has been included here for clarity and completion.}
    \label{fig:enter-label}
\end{figure}

\renewcommand{\arraystretch}{1.5}
\begin{table}[h]
\centering

\begin{tabular}{ccccccc}
\hline
\hline
\textbf{Atom} & \textbf{Site} & \textit{x} & \textit{y} & \textit{z} & \textbf{B$_{\textnormal{iso}}$} & \textbf{Occupancy} \\
\hline
Bi1 & 8c & 0.95685(69) & 0.23741(24) & 0.14649(38) & 0.130(66) & 1 \\
Yb & 8c & 0.00808(65) & 0.05396(15) & 0.35944(30) & 0.042(50) & 0.808(26) \\
Bi2 & 8c & 0.00808(65) & 0.05396(15) & 0.35944(30) & 0.042(50) & 0.192(26) \\
Ge & 8c & 0.00543(93) & 0.40456(27) & 0.40269(34) & 0.282(62) & 1 \\
O1 & 8c & 0.07054(97) & 0.30183(32) & 0.32959(48) & 0.21(11) & 1 \\
O2 & 8c & 0.3002(11) & 0.43740(39) & 0.45670(53) & 0.19(8) & 1 \\
O3 & 8c & 0.3182(14) & 0.12552(40) & 0.47399(56) & 0.62(13) & 1 \\
O4 & 8c & 0.2542(13) & 0.15041(43) & 0.23658(57) & 0.50(10) & 1 \\
O5 & 8c & 0.3493(11) & 0.47814(43) & 0.19261(54) & 0.37(10) & 1 \\
\hline
\hline
\end{tabular}

\caption{The refined parameters of atomic coordinates \textit{(x, y, z)}, isotropic atomic displacement \textit{(B$_\textnormal{iso}$)}, and site occupancy obtained by the Rietveld refinement of the neutron diffraction pattern of \bygo\ taken at 20 K using $\lambda$ = 1.54 \AA\ incident wavelength.\label{table_refinement}}
\end{table}

\begin{figure}[h]
    \centering
    \includegraphics[width=1\linewidth]{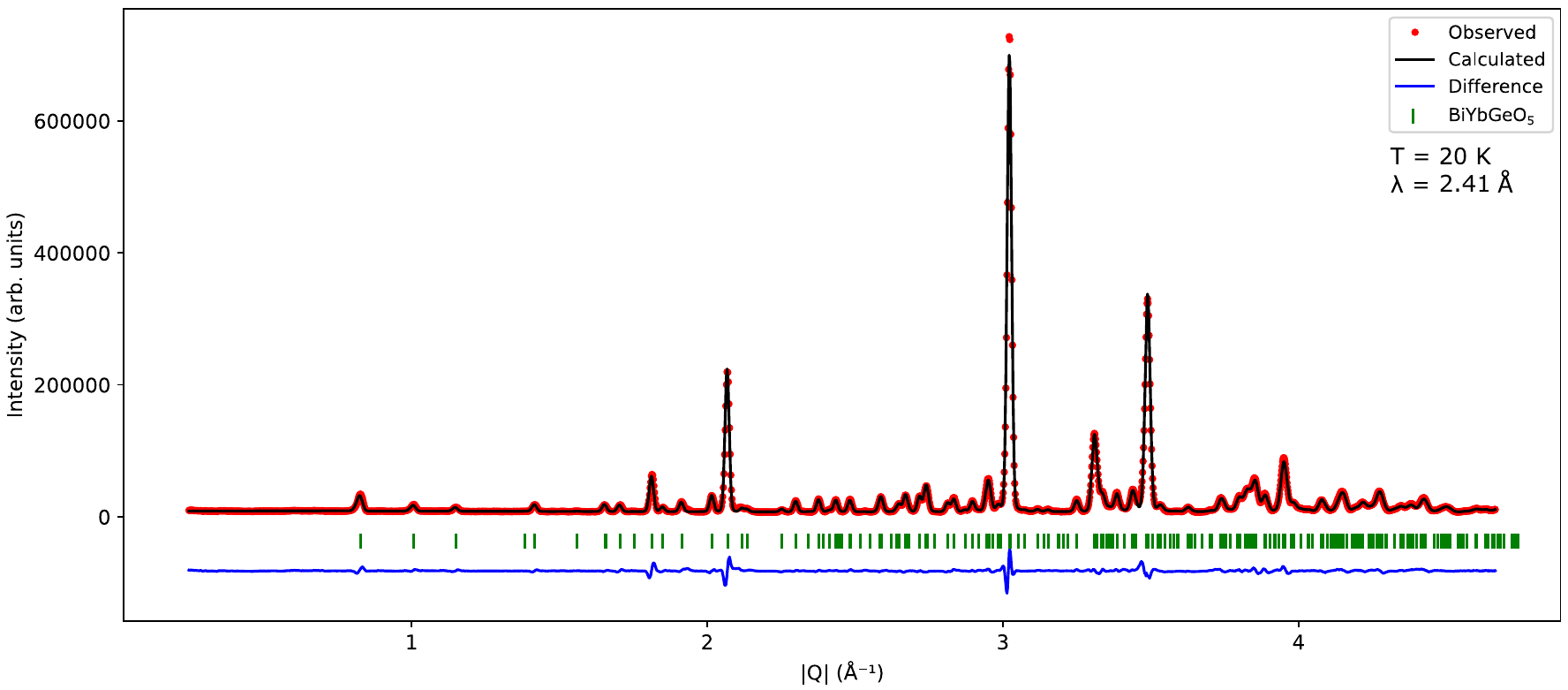}
    \caption{A Rietveld refinement of the structure of \bygo\ obtained from neutron diffraction performed at 20 K using an incident wavelength of $\lambda$ = 2.41 \AA. The peaks at 3.02 \AA and 3.49 \AA, come from the Cu canister used for the measurement. These Cu peaks were fitted using Le Bail extraction, not to influence the refinement of the \bygo\ parameters. The \textnormal{$R_{bragg}$} for the pattern from \bygo\ is 4.407 ($\chi^2 = 32.48$).}
    \label{fig:enter-label}
\end{figure}

\renewcommand{\arraystretch}{1.5}
\begin{table}[h]
\centering

\begin{tabular}{ccccccc}
\hline
\hline
\textbf{Atom} & \textbf{Site} & \textit{x} & \textit{y} & \textit{z} & \textbf{B$_{\textnormal{iso}}$} & \textbf{Occupancy} \\
\hline
Bi1 & 8c & 0.96054(86) & 0.23653(30) & 0.14544(46) & 0.130 & 1 \\
Yb1 & 8c & 0.00755(78) & 0.05422(17) & 0.35863(41) & 0.042 & 0.810(20) \\
Bi2 & 8c & 0.00755(78) & 0.05422(17) & 0.35863(41) & 0.042 & 0.190(20) \\
Ge1 & 8c & 0.0112(13) & 0.40523(30) & 0.40316(37) & 0.282 & 1 \\
O1 & 8c & 0.0672(11) & 0.30067(42) & 0.33123(60) & 0.21 & 1 \\
O2 & 8c & 0.2981(15) & 0.43896(42) & 0.45953(73) & 0.19 & 1 \\
O3 & 8c & 0.3089(15) & 0.12604(46) & 0.47127(57) & 0.62 & 1 \\
O4 & 8c & 0.2543(18) & 0.15153(48) & 0.23512(65) & 0.50 & 1 \\
O5 & 8c & 0.3542(13) & 0.47883(57) & 0.19461(63) & 0.37 & 1 \\
\hline
\hline
\end{tabular}

\caption{The refined parameters of atomic coordinates \textit{(x, y, z)}, and site occupancy obtained by the Rietveld refinement of the neutron diffraction pattern of \bygo\ taken at 20 K using $\lambda$ = 2.41 \AA\ incident wavelength.}
\end{table}

\begin{figure}[h]
    \centering
    \includegraphics[width=1\linewidth]{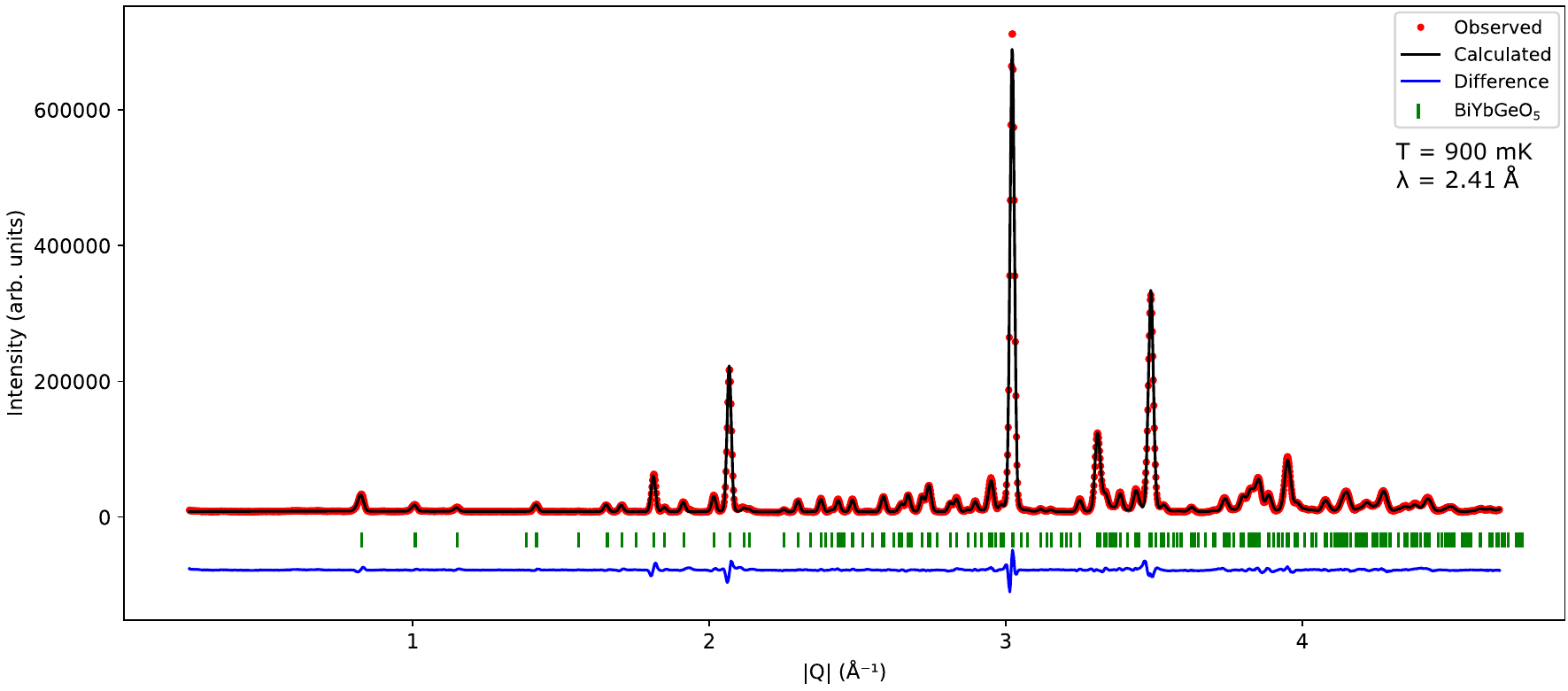}
    \caption{A Rietveld refinement of the structure of \bygo\ obtained from neutron diffraction performed at 900 mK using an incident wavelength of $\lambda$ = 2.41 \AA. The peaks at 3.02 \AA and 3.49 \AA, come from the Cu canister used for the measurement. These Cu peaks were fitted using Le Bail extraction, not to influence the refinement of the \bygo\ parameters. The \textnormal{$R_{bragg}$} for the pattern from \bygo\ is 4.412 ($\chi^2 = 26.66$).}
    \label{fig:enter-label}
\end{figure}

\renewcommand{\arraystretch}{1.5}
\begin{table}[h]
\centering

\begin{tabular}{ccccccc}
\hline
\hline
\textbf{Atom} & \textbf{Site} & \textit{x} & \textit{y} & \textit{z} & \textbf{B$_{\textnormal{iso}}$} & \textbf{Occupancy} \\
\hline
Bi1 & 8c & 0.95967(77) & 0.23645(27) & 0.14565(41) & 0.130 & 1 \\
Yb & 8c & 0.00872(71) & 0.05410(16) & 0.35890(38) &  0.042 & 0.830(18) \\
Bi2 & 8c & 0.00872(71) & 0.05410(16) & 0.35890(38) & 0.042 & 0.170(18) \\
Ge & 8c & 0.00911(12) & 0.40488(27) & 0.40311(33) & 0.282 & 1 \\
O1 & 8c & 0.0663(10) & 0.30116(38) & 0.33148(54) & 0.21 &  1 \\
O2 & 8c & 0.2962(14) & 0.43925(38) & 0.45929(66) & 0.19  &  1 \\
O3 & 8c & 0.3090(14) & 0.12623(42) & 0.47163(52) & 0.62 &  1 \\
O4 & 8c & 0.2536(16) & 0.15194(44) & 0.23521(59) & 0.50 &  1 \\
O5 & 8c & 0.3557(12) & 0.47913(51) & 0.19362(57) & 0.37 &  1 \\
\hline
\hline
\end{tabular}

\caption{The refined parameters of atomic coordinates \textit{(x, y, z)}, and site occupancy obtained by the Rietveld refinement of the neutron diffraction pattern of \bygo\ taken at 900 mK using $\lambda$ = 2.41 \AA\ incident wavelength.}
\end{table}

\begin{figure}[h]
    \centering
    \includegraphics[width=1\linewidth]{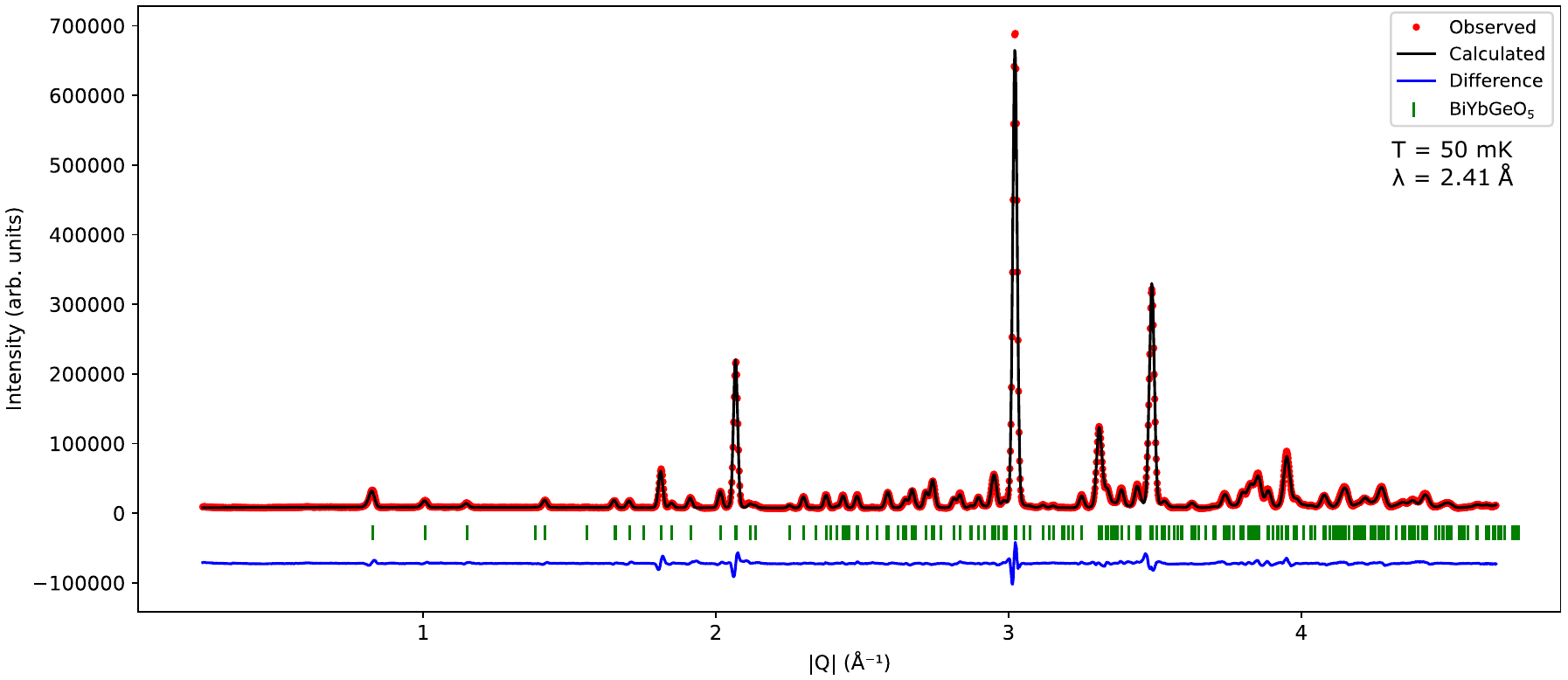}
    \caption{A Rietveld refinement of the structure of \bygo\ obtained from neutron diffraction performed at 50 mK using an incident wavelength of $\lambda$ = 2.41 \AA. The peaks at 3.02 \AA and 3.49 \AA, come from the Cu canister used for the measurement. These Cu peaks were fitted using Le Bail extraction, not to influence the refinement of the \bygo\ parameters. The \textnormal{$R_{bragg}$} for the pattern from \bygo\ is 4.538($\chi^2=28.92$).} 
    \label{fig:enter-label}
\end{figure}

\renewcommand{\arraystretch}{1.5}
\begin{table}[h]
\centering
\begin{tabular}{ccccccc}
\hline
\hline
\textbf{Atom} & \textbf{Site} & \textit{x} & \textit{y} & \textit{z} & \textbf{B$_{\textnormal{iso}}$} & \textbf{Occupancy} \\
\hline
Bi1 & 8c & 0.96074(82) & 0.23671(28) & 0.14583(43) & 0.130 & 1 \\
Yb & 8c & 0.00843(74) & 0.05420(17) & 0.35896(40) & 0.042 & 0.821(22) \\
Bi2 & 8c & 0.00843(74) & 0.05420(17) & 0.35896(40) & 0.042 & 0.179(22) \\
Ge1 & 8c & 0.0112(13) & 0.40493(29) & 0.40286(35) &  0.282 & 1 \\
O1 & 8c & 0.0672(10) & 0.30081(40) & 0.33139(57) & 0.21 & 1 \\
O2 & 8c & 0.2974(14) & 0.43953(40) & 0.46021(70) & 0.19  & 1 \\
O3 & 8c & 0.3087(14) & 0.12620(45) & 0.47118(55) & 0.62 & 1 \\
O4 & 8c & 0.2544(17) & 0.15143(46) & 0.23524(63) & 0.50 & 1 \\
O5 & 8c & 0.3548(12) & 0.47905(54) & 0.19409(60) & 0.37 & 1 \\
\hline
\hline
\end{tabular}
\caption{The refined parameters of atomic coordinates \textit{(x, y, z)}, and site occupancy obtained by the Rietveld refinement of the neutron diffraction pattern of \bygo\ taken at 50 mK using $\lambda$ = 2.41 \AA\ incident wavelength.}
\end{table}


\newpage

\section{INELASTIC NEUTRON SCATTERING}

\subsection{$\Delta$E vs |Q| Full Range Slice}
\begin{figure}[h]
    \centering
    \includegraphics[width=1\linewidth]{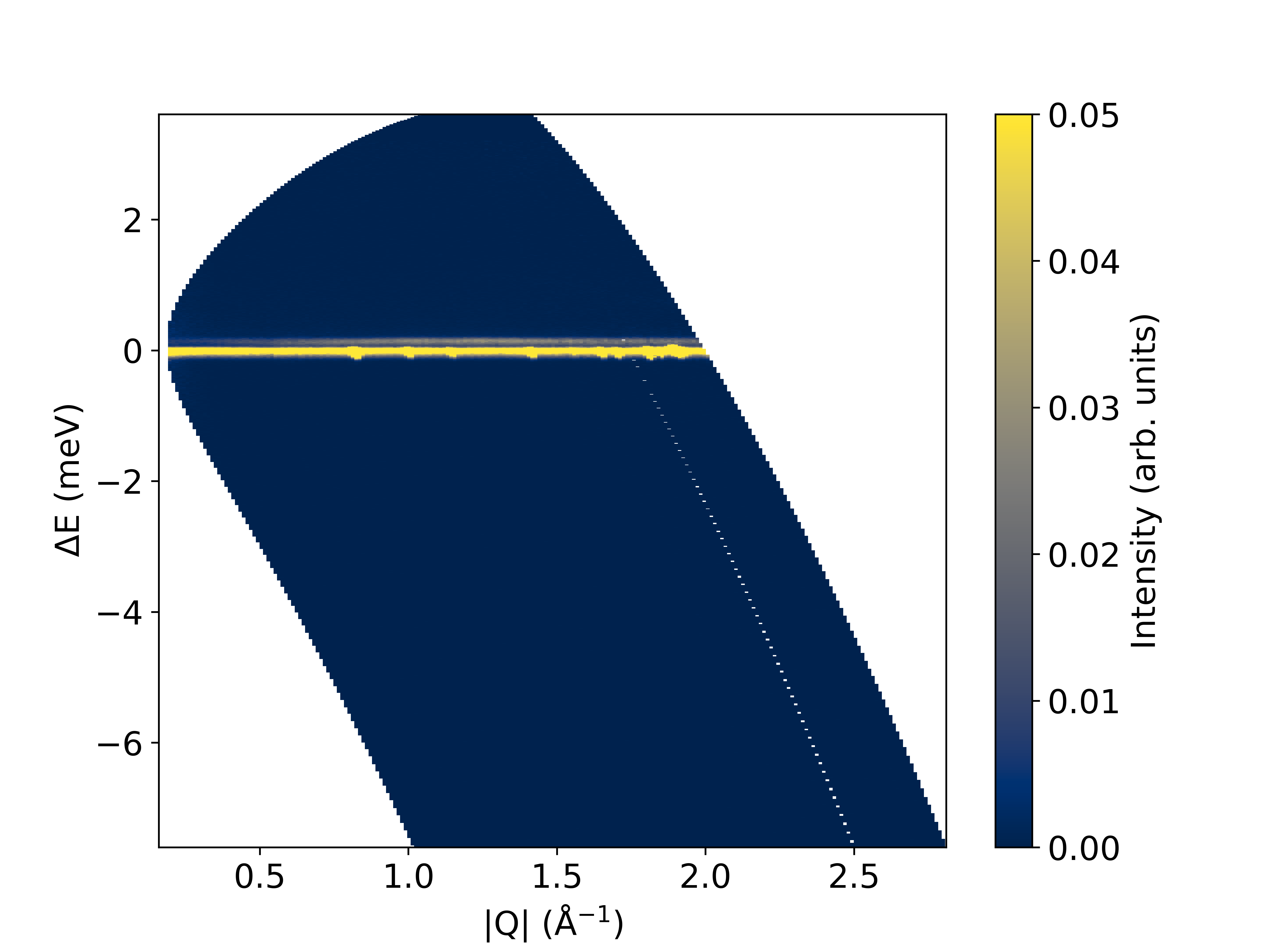}
    \caption{$\Delta$E vs |Q| color contour slice of the full range of the INS data collected at 250 mK. No other features are visible except for the bright elastic line and the faint flat triplon band.}
    \label{hyspec_full_slice}
\end{figure}

Figure~\ref{hyspec_full_slice} shows the full range of the INS data collected at 250 mK. It shows a bright band at the elastic line and a flat triplon band. The energy transfer range goes from -7.59 meV to 3.61 meV, in which no crystalline electric field (CEF) excitations were detected, placing a lower bound on the CEF gap of 42 K.

\subsection{Evolution of the Triplon Excitation with Temperature}
\begin{figure}[h]
    \centering
    \includegraphics[width=1\linewidth]{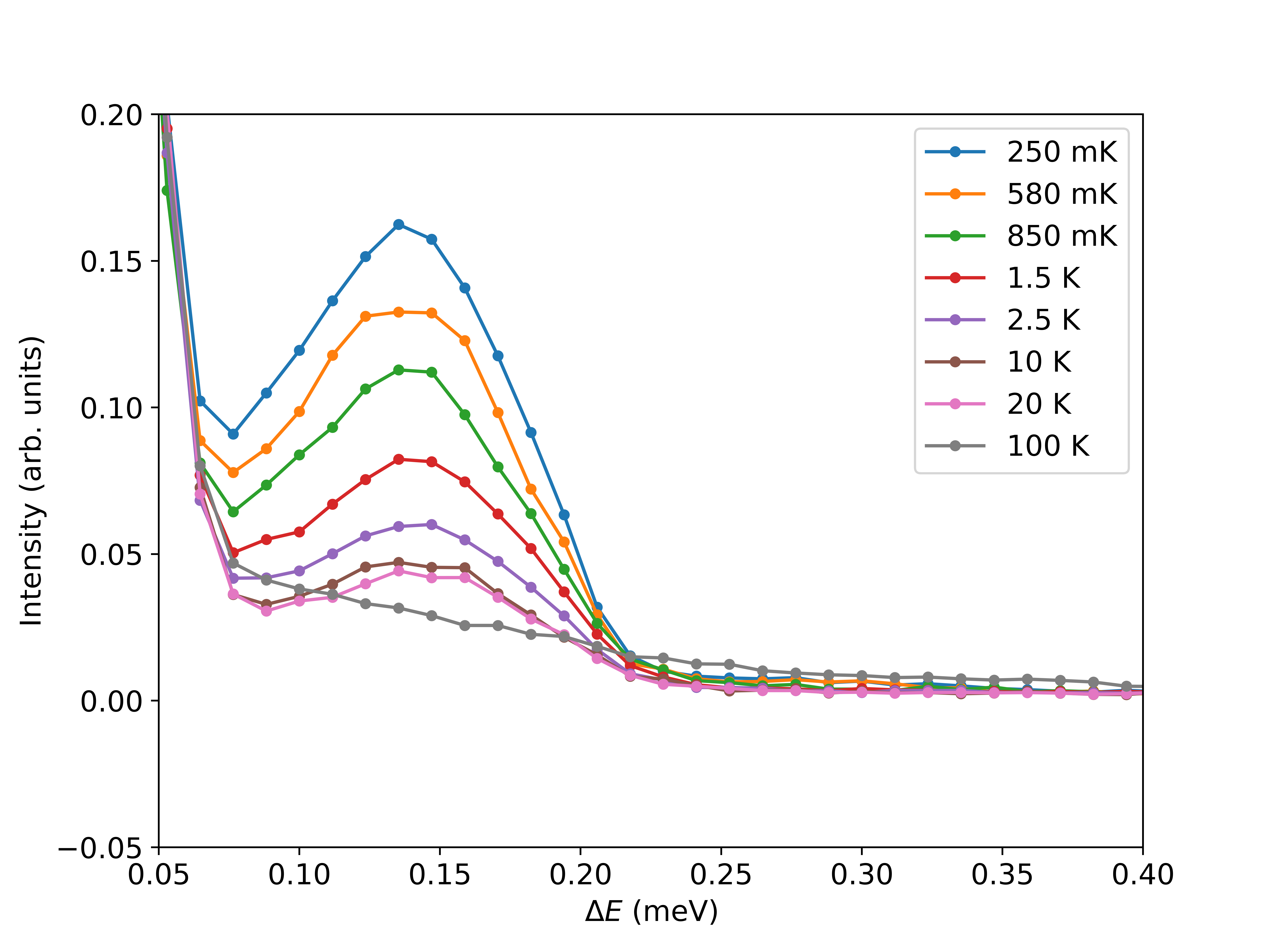}
    \caption{Cuts of the INS data made along $\Delta$E integrating |Q| from 0.25 \AA$^{-1}$ to 1.4 \AA$^{-1}$ for 250 mK, 300 mK, 580 mK, 850 mK, 1.5 K, 2.5 K, 10 K, and 20 K, and 100 K. The intensities of all scans have been normalized to allow comparison between the spectral weights of the triplon peak as a function of temperature. A temperature-dependent triplon mode centered at 0.135 meV is visible for all three temperatures.}
    \label{triplon_vs_temperature}
\end{figure}

Although the intensity decreases as temperature increases, the width of the excitation stays the same. Figure~\ref{triplon_vs_temperature} shows the evolution of the triplon across all of the temperatures that were used to measure the inelastic spectra.

\subsection{Gaussian Fits of I($\Delta$E) Cuts of Inelastic Neutron Scattering}

Gaussians were given by:

$$I(\Delta E)=H + ae^{-\left( \frac{(\Delta E -b)^2}{2c^2}\right)}$$

where, \\
$H$ is the baseline for the curve \\
$a$ is the amplitude of the Gaussian \\
$b$ is the center of the Gaussian \\
$c$ is the standard deviation or the Gaussian RMS  

FWHM = $\sqrt{2ln2}\cdot c \approx 2.355 c$ \\

Mohanty \textit{et al.} \cite{mohanty_disordered_2023} used heat capacity, which is especially susceptible to energy hierarchies, to determine that the dimers in BiYbGeO$_5$ have an XXZ-type anisotropy. Using this hypothesis, we used the inelastic spectrum collected on the material to see what the exchange parameters would come out as if XXZ-type anisotropy were considered. For the sake of completion, single and triple Gaussian fits were also made, which resulted in comparable fit qualities.

\begin{figure}[h]
    \centering
    \includegraphics[width=1\linewidth]{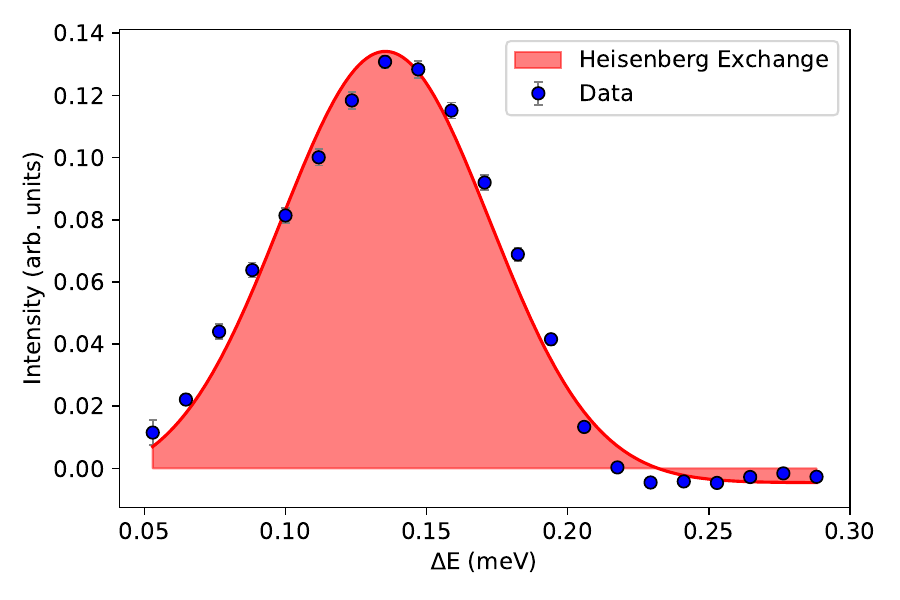}
    \caption{A cut along $\Delta$E made by subtracting the 100 K INS data from the 250 mK data. This excitation has been fit to a Gaussian function, centered at 0.1354(4) meV. The full width at half maximum for the Gaussian is 0.09 meV, which is comparable to the resolution of HYSPEC ($\sim$ 0.1 meV).}
    \label{single_gaussian_fit}
\end{figure}

A single Gaussian fit would suggest a Heisenberg interaction where the triplet excitation is triply degenerate. As the resolution of the instrument limits the triplon excitation observed, attempts to fit the excitation to a single-Gaussian model yielded a full-width at half-maximum (FWHM) comparable to the instrument's resolution (see Figure~\ref{single_gaussian_fit}). This fit has a high $\chi^2$ value of 229.449, indicating a poor fit. The fitted parameters are given in Table~\ref{single_gaussian_parameters}. \\

\begin{figure}[h]
    \centering
    \includegraphics[width=1\linewidth]{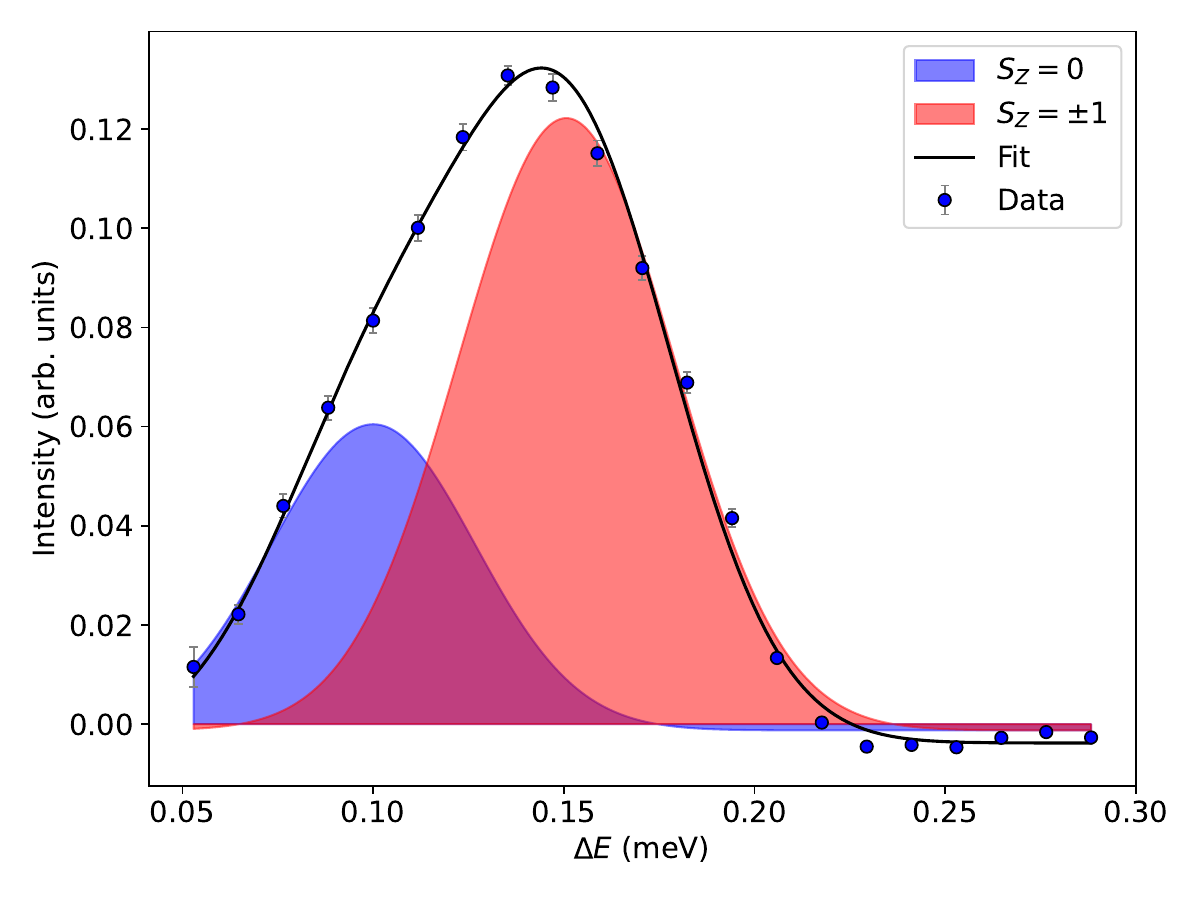}
    \caption{A cut along $\Delta$E made by subtracting the 100 K INS data from the 250 mK. This excitation has been fit to the sum of two Gaussians, centered at 0.100(1) meV and 0.1510(4) meV, consistent with an XXZ-type intradimer exchange interaction. The full widths at half maximum for the Gaussians are 0.06 meV and 0.07 meV, respectively, which are smaller than the resolution of HYSPEC ($\sim$ 0.1 meV).This is the same as Figure 4 in the main text and has been included here for clarity and completion.}
    \label{double_gaussian)fit}
\end{figure}

A double Gaussian model would suggest an XXZ interaction such that $S_z=\pm1$ states are doubly degenerate and $S_z=0$ has a separate energy. The spectral weight of the doubly degenerate state should be double that of the separate energy, and that was taken as a constraint while fitting the double Gaussian. This fit has a  $\chi^2$ value of 68.429 (see Figure~\ref{double_gaussian)fit}). The fitted parameters are given in Table~\ref{double_gaussian_parameters}. \\

\begin{figure}[h]
    \centering
    \includegraphics[width=1\linewidth]{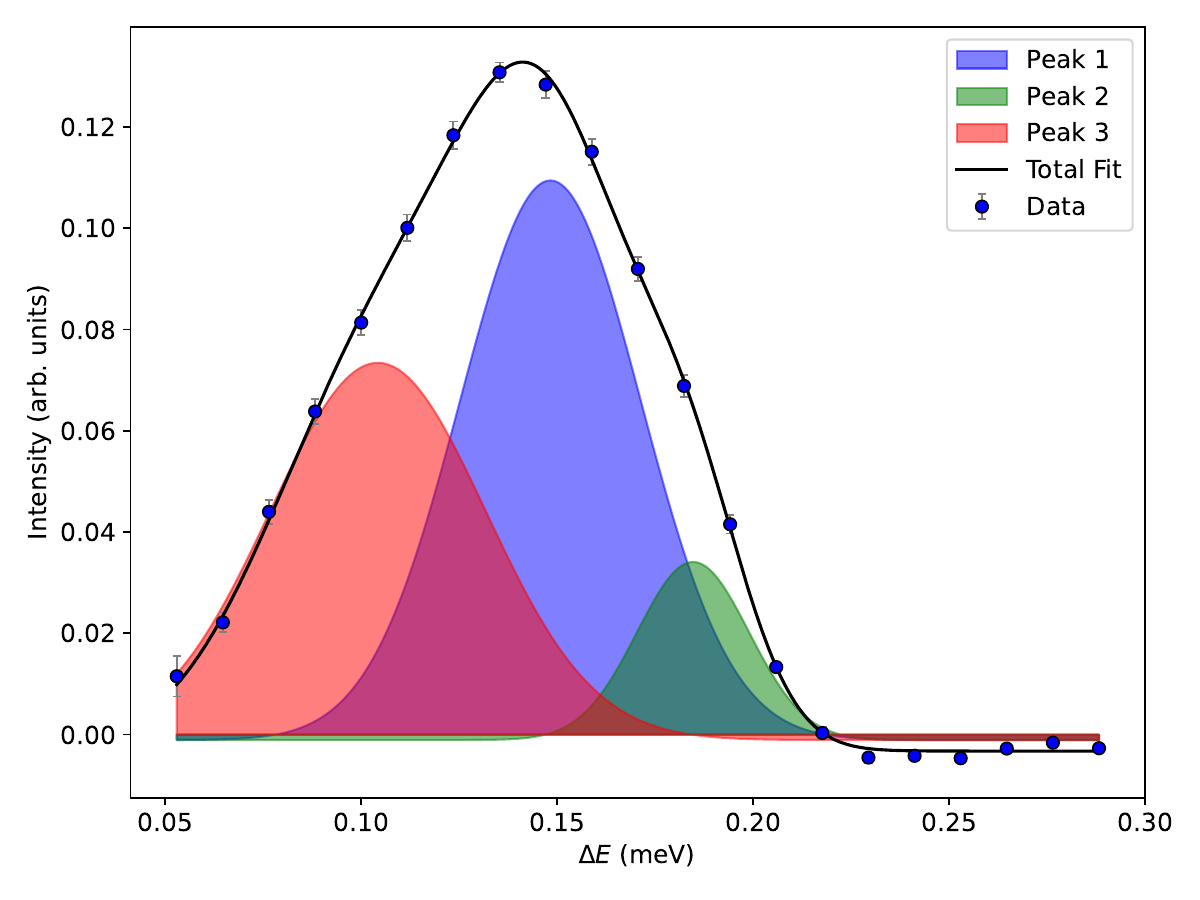}
    \caption{A cut along $\Delta$E made by subtracting the 100 K INS data from the 250 mK. This excitation has been fit to the sum of three Gaussians centered at 0.104(23) meV, 0.148(7) meV, and 0.185(5) meV. The full widths at half maximum for the Gaussians are 0.06 meV, 0.05 meV, and 0.03 meV, respectively, which are smaller than the resolution of HYSPEC ($\sim$ 0.1 meV).}
    \label{triple_gaussian_fit}
\end{figure}

A triple Gaussian model would mean an XYZ interaction such that $S_z=+1$, $S_z=-1$, and $S_z=0$ all have different energies. This leads to a better fit, as expected, based on the increase in the number of refined parameters (see Figure~\ref{triple_gaussian_fit}). The fitted parameters are given in Table~\ref{triple_gaussian_parameters}.

\subsection{Structure Factor Fits for Dimer Distance}
\begin{figure}[h]
    \centering
    \includegraphics[width=1\linewidth]{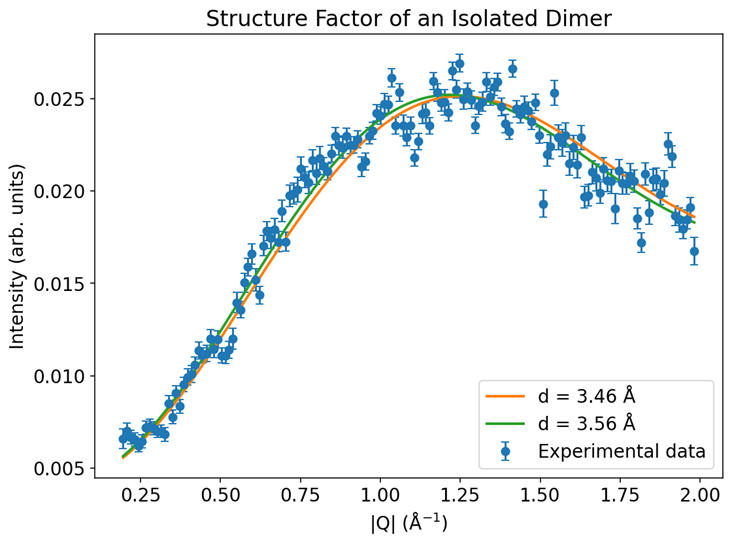}
    \caption{A cut of the INS data taken at 250 mK made along |Q| by integrating $\Delta$E symmetrically around the 0.135 meV maximum from 0.100 meV to 0.170 meV. The two lines are the simulated structure factor of BiYbGeO$_5$ if the dimer distance were taken to be 3.46 \AA\ (nearest neighbor distance) or 3.56 \AA\ (next nearest neighbor distance).}
    \label{structure_factor_comparision}
\end{figure}

The presence of Bi atoms on Yb sites disrupts a fraction of the dimers, but does not alter the form of the structure factor itself. The primary effect is a reduction in overall intensity, reflecting the smaller number of intact dimers in the system. The intradimer distance extracted from the structure factor should remain unchanged, as it is determined by the geometry of the lattice. However, it is difficult to make definitive statements about possible dimerization involving next-nearest-neighbor Yb ions in the presence of broken nearest-neighbor dimers, since the NN and NNN distances are too close to be distinguished within the resolution of the INS data. Figure~\ref{structure_factor_comparision} shows how close the fits to the structure factor would be if the dimer distance were taken to be the NN or the NNN distance. Future single-crystal INS measurements, with improved Q-resolution, could provide a more sensitive probe of whether alternative dimer configurations emerge in the presence of site disorder.

\renewcommand{\arraystretch}{1.5}
\begin{table}[h]
    \centering
    \begin{tabular}{c|c}
    \hline
    \hline
        \textbf{Parameter} & \textbf{Gaussian} \\
        \hline
        H & $-0.004601 \pm 0.000371$\\
        \hline
        a & $0.13876 \pm 0.00121$\\
        \hline
        b & $0.135427 \pm 0.000361$\\
        \hline
        c & $0.037026 \pm 0.000344$\\
        \hline
        FWHM & $\approx$ 0.09\\
        \hline
        $\chi^2$ & 229.449\\
    \hline
    \hline
    \end{tabular}
    \caption{Parameters for single Gaussian fit.}
    \label{single_gaussian_parameters}
\end{table}

\renewcommand{\arraystretch}{1.5}
\begin{table}[h]
    \centering
    \begin{tabular}{c|c|c}
    \hline
    \hline
        \textbf{Parameter} & \textbf{Gaussian 1} & \textbf{Gaussian 2} \\
        \hline
        H & \multicolumn{2}{c}{$-0.00381 \pm 0.000353$} \\
        \hline
        a & $0.0617 \pm 0.00122$ & $0.123 \pm 0.00243$\\
        \hline
        b & $0.100 \pm 0.001$ & $0.1510 \pm 0.0004$\\
        \hline
        c & $0.0268 \pm 0.00127$ & $0.0284 \pm 0.000416$\\
        \hline
        FWHM & $\approx$ 0.06 & $\approx$ 0.07\\
        \hline
        $\chi^2$ & \multicolumn{2}{c}{68.429}\\
    \hline
    \hline
    \end{tabular}
    \caption{Parameters for double Gaussian fit.}
    \label{double_gaussian_parameters}
\end{table}

\renewcommand{\arraystretch}{1.5}
\begin{table}[h]
    \centering
    \begin{tabular}{c|c|c|c}
    \hline
    \hline
        \textbf{Parameter} & \textbf{Gaussian 1} & \textbf{Gaussian 2} & \textbf{Gaussian 3} \\
        \hline
        H & \multicolumn{3}{c}{$-0.003258 \pm 0.00035$} \\
        \hline
        a & $0.111 \pm 0.0608$ & $0.0745 \pm 0.0451$ & $0.0352 \pm 0.0323$ \\
        \hline
        b & $0.148 \pm 0.00666$ & $0.104 \pm 0.0231$ & $0.185 \pm 0.00481$ \\
        \hline
        c & $0.0231 \pm 0.00928$ & $0.0275 \pm 0.0088$ & $0.0143 \pm 0.00377$ \\
        \hline
        FWHM & $\approx$ 0.05 & $\approx$ 0.06 & $\approx$ 0.03 \\
        \hline
        $\chi^2$ & \multicolumn{3}{c}{14.168} \\
    \hline
    \hline
    \end{tabular}
    \caption{Parameters for triple Gaussian fit.}
    \label{triple_gaussian_parameters}
\end{table}

\bibliography{bygo}